\documentclass[a4paper,fleqn,usenatbib]{mnras}
\usepackage[normalem]{ulem}

\usepackage{graphicx}	
\usepackage{amsmath}	
\usepackage{amssymb}	

\defcitealias{2015AJ....149...91P}{Paper\,I}
\defcitealias{2015MNRAS.447..927M}{Paper\,II}
\defcitealias{2015MNRAS.451..312N}{Paper\,IV}
\defcitealias{2017MNRAS.464.3636M}{Paper\,IX}
\defcitealias{2016MNRAS.463..449S}{Paper\,X}
\defcitealias{2015MNRAS.454.4197R}{Paper\,V}

\title[Multiple stellar populations in M\,15]{The {\it Hubble Space Telescope} UV Legacy Survey of Galactic
 Globular Clusters - XIV. Multiple stellar populations within M\,15 and their radial distribution.\thanks{Based on observations with the NASA/ESA
{\it Hubble Space Telescope}, obtained at the Space Telescope Science Institute, which is operated by
AURA, Inc., under NASA contract NAS 5-26555.}}

\author[Nardiello et al.]{D.\ Nardiello$^{1,2}$\thanks{E-mail: domenico.nardiello@unipd.it}, 
A.\,P.\, Milone$^{1}$,
G.\,Piotto$^{1,2}$, 
J.\,Anderson$^{3}$,
L.\,R.\, Bedin$^{2}$,
A.\,Bellini$^{3}$,
\newauthor
S.\,Cassisi$^{4}$,
M.\ Libralato$^{3}$, 
and A.\,F.\, Marino$^{5}$ \\
$^{1}$Dipartimento di Fisica e Astronomia ``Galileo Galilei'', Universit\`a di Padova, Vicolo dell'Osservatorio 3, I-35122, Padova, Italy \\
$^{2}$Istituto Nazionale di Astrofisica - Osservatorio Astronomico di Padova, Vicolo dell'Osservatorio 5, I-35122, Padova, Italy \\
$^{3}$Space Telescope Science Institute, 3800 San Martin Drive, Baltimore, MD 21218, USA \\
$^{4}$Osservatorio Astronomico d'Abruzzo, Via M. Maggini sn., I-64100 Teramo, Italy \\
$^{5}$Research School of Astronomy and Astrophysics, The Australian National University, Cotter Road, Weston, ACT, 2611, Australia \\
}

\date{Accepted 2018 March 14. Received 2018 March 13; in original form 2018 February 22}

\pubyear{2017}

\begin{document}
\label{firstpage}
\pagerange{\pageref{firstpage}--\pageref{lastpage}}
\maketitle

\begin{abstract}
   In the context of the {\it Hubble Space Telescope} UV Survey of
   Galactic Globular Clusters (GCs), we derived high-precision,
   multi-band photometry to investigate the multiple stellar
   populations in the massive and metal-poor GC M\,15.  By creating for red-giant branch (RGB) stars of
   the cluster a `chromosome map', which is a pseudo two-colour
   diagram made with appropriate combination of F275W, F336W, F438W,
   and F814W magnitudes, we revealed colour spreads
     around two of the three already known stellar populations. These
     spreads cannot be produced by photometric errors alone and could
     hide the existence of (two) additional populations.  This
   discovery increases the complexity of the multiple-population
   phenomenon in M\,15.

    Our analysis shows that M\,15 exhibits a faint sub-giant branch
    (SGB), which is also detected in colour-magnitude diagrams (CMDs)
    made with optical magnitudes only. This poorly-populated SGB
    includes about 5\% of the total number of SGB stars and evolves
    into a red RGB in the $m_{\rm F336W}$ vs.\,$m_{\rm F336W}-m_{\rm
      F814W}$ CMD, suggesting that M\,15 belongs to the class of
    Type\,II GCs.

    We measured the relative number of stars in each population at
    various radial distances from the cluster centre, showing that all
    of these populations share the same radial distribution within
    statistic uncertainties. These new findings are discussed in the
    context of the formation and evolution scenarios of the multiple
    populations.
  
\end{abstract}




\section{Introduction}
\label{sec:intro}
In the last few years, several scenarios for the formation and
  evolution of multiple stellar populations (MPs) in globular clusters
  (GCs) have been suggested.  Some authors claim that GCs
  host a primordial first stellar generation (1G), and a second generation of
  stars (2G) formed from matter ejected by polluters belonging to the 1G
  (e.g.\,\citealt{2001ApJ...550L..65V,2007A&A...464.1029D,2016MNRAS.458.2122D}).

An alternative possibility is that GCs host only a single generation
of stars, and the distinct populations of stars with different
abundance of helium and light elements would be the product of stellar
interactions in the unique dense environment of proto GCs
(e.g.\,\citealt{2009A&A...507L...1D,2013MNRAS.436.2398B}; 
  \citealt{2014MNRAS.437L..21D}, see \citealt{2015MNRAS.454.4197R},
  hereafter \citetalias{2015MNRAS.454.4197R}, for a critical
  discussion).

The {\it Hubble Space Telescope UV Legacy Survey of Galactic GCs}
(GO-13297, PI:~G.\,Piotto; \citetalias{2015AJ....149...91P}), is a
{\it HST} treasury project aimed at discriminating among these scenarios
and constraining the formation and evolution of MPs in GCs.
Specifically, we collected F275W, F336W, and F438W images of 57 GCs
previously observed in F606W and F814W bands (GO-10775,
PI:~A.\,Sarajedini, see \citealt{2007AJ....133.1658S}), to
characterise for the first time MPs in a large sample of
clusters. This dataset is complemented by data previously collected from GO-12311
(PI:~G.\,Piotto) and GO-12605 (PI:~G.\,Piotto), which were
pilot projects for GO-13297.

Results from this program have already provided a major breakthrough towards
understanding the MP phenomenon.  We have detected two or more
populations in all the analysed GCs, thus suggesting that MPs are
indeed ubiquitous in Galactic GCs (\citetalias{2015AJ....149...91P}). The
number of distinct populations, their chemical compositions, and the
relative fractions of 2G and 1G stars dramatically change from one
cluster to another, depending on the mass of the host GC. This fact
suggests that cluster mass has played a major role to determine the
MP phenomenon (\citealt{2010A&A...516A..55C}; \citealt{2017MNRAS.464.3636M}, hereafter
\citetalias{2017MNRAS.464.3636M}).

Moreover, this unique dataset allowed us to investigate the relative
ages and the internal kinematics of the distinct populations
(\citealt{2015MNRAS.451..312N,2015ApJ...810L..13B}).

Strong constraints on formation scenarios could be provided by the
spatial distribution of MPs. Indeed, it has been suggested that 2G
stars form in the innermost cluster region via cooling flow
(e.g.\,\citealt{2008MNRAS.391..825D,2010MNRAS.407..854D}).  The fact
that some clusters would still retain information on the initial
distribution of 1G and 2G stars makes the study of their radial
distribution a powerful tool to shed light on the MP phenomenon (
  \citealt{2008AN....329..976D,2013MNRAS.429.1913V}).

In this context, NGC\,7078 (M\,15) is an intriguing case.  The cluster
M\,15 is a massive ($\sim 5.6\time10^5 M_\odot$,
\citealt{2010MNRAS.406.2000M};  \citealt{2017MNRAS.471.3668S})
metal-poor Galatic GC ([Fe/H]$\sim -2.37$, \citealt[updated to
  December 2010]{1996AJ....112.1487H};  \citealt{2011AJ....141..175S}), characterized by many chemical peculiarities (see, e.g., \citealt{2000JKAS...33..137L, 2005AJ....130.1177C, 2009A&A...505..117C, 2010A&A...524A..44P}).  It has been suggested that in
the $\sim 2 \times 2$ arcmin$^2$ inner region of this cluster, 1G
stars are more centrally concentrated than 2G stars
(\citealt{2015ApJ...804...71L}) in contrast with what is observed in
other GCs
(e.g.\,\citealt{2007ApJ...654..915S,2009A&A...507.1393B,2012ApJ...744...58M,2013ApJ...765...32B,2016MNRAS.463..449S},
hereafter \citetalias{2016MNRAS.463..449S}). At radial distances
larger than $\sim$2 arcmin, the radial trend is the opposite and 2G
stars are more centrally concentrated than 1G stars
(\citealt{2011A&A...525A.114L}).  Such an unusual, ``U-shaped''
spatial distribution of 1G and 2G stars, if real, would be a major
challenge for the scenarios of formation and evolution of MPs.

In this study, we exploit {\it HST} multi-band photometry to investigate
the MPs within M\,15 and derive their radial distribution. The paper
is organised as follows. In Section~\ref{sec:obs} we describe the
observations and the data reduction. We identify the multiple
populations in Section~\ref{sec:chrm}, while in
Section~\ref{sect:frac} we measure the relative fractions of stars in
each population. The radial distribution of the distinct stellar
populations is derived in Section~\ref{sec:raddis}. Summary and
discussion are provided in Section~\ref{sec:summ}.

\section{Observations and data reduction}
\label{sec:obs}

\begin{table*}
  \caption{Description of the archive {\it HST} images reduced in this analysis.}
    \label{tab1}
    \begin{tabular}{ c c c c c c}
\hline
\multicolumn{1}{c}{Program} &
\multicolumn{1}{c}{Epoch} &
\multicolumn{1}{c}{Filter} &
\multicolumn{1}{c}{$N \times $ Exp.~time} &
\multicolumn{1}{c}{Instrument} &
\multicolumn{1}{c}{PI} \\ 
\hline

10775 &  2006.33 & F606W &  15\,s $+$ 4 $\times$ 130\,s & ACS/WFC &  A.~Sarajedini \\
10775 &  2006.33 & F814W &  15\,s $+$ 4 $\times$ 150\,s & ACS/WFC &  A.~Sarajedini \\
12605 &  2011.80 & F275W &  2 $\times$ 615\,s $+$ 3$\times$ 700\,s & WFC3/UVIS & G.~Piotto\\
12605 &  2011.80 & F336W &  5 $\times$ 350\,s & WFC3/UVIS & G.~Piotto\\
12605 &  2011.80 & F438W &  5 $\times$ 65\,s  & WFC3/UVIS & G.~Piotto\\
13295 &  2013.67 & F343N &  2 $\times$ 350\,s & WFC3/UVIS & S.S.~Larsen \\
13295 &  2013.67 & F555W &  2 $\times$  10\,s & WFC3/UVIS & S.S.~Larsen \\
\hline
\end{tabular}

\end{table*}

\begin{figure}
\includegraphics[bb=20 216 560 700, width=0.5\textwidth]{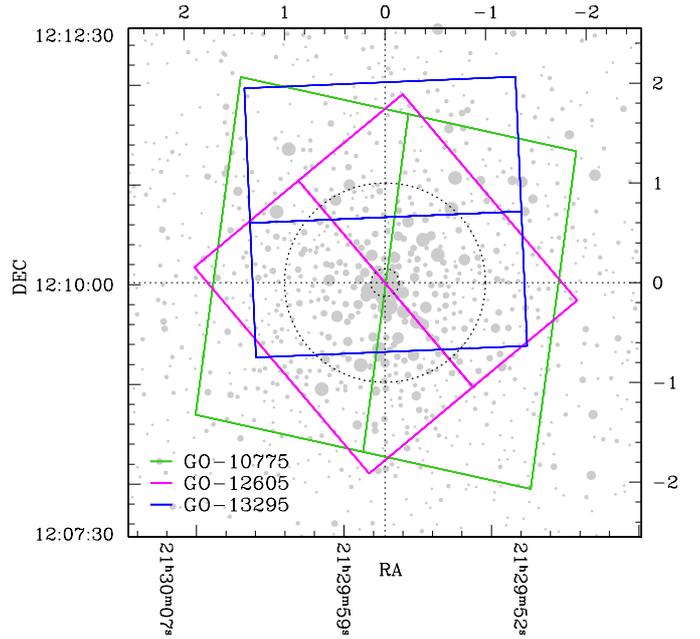}
\caption{Footprint of the {\it HST} ACS and WFC3 observations used in
  this analysis. The inner circle is the core radius ($r_c
  =0.14$\,arcmin), the outer circle is the half-light radius
  ($r_h=1.0$\,arcmin; \citealt[updated to December
    2010]{1996AJ....112.1487H}). \label{fig1}}
\end{figure}

\begin{figure*}
\includegraphics[bb=20 232 584 571, width=0.95\textwidth]{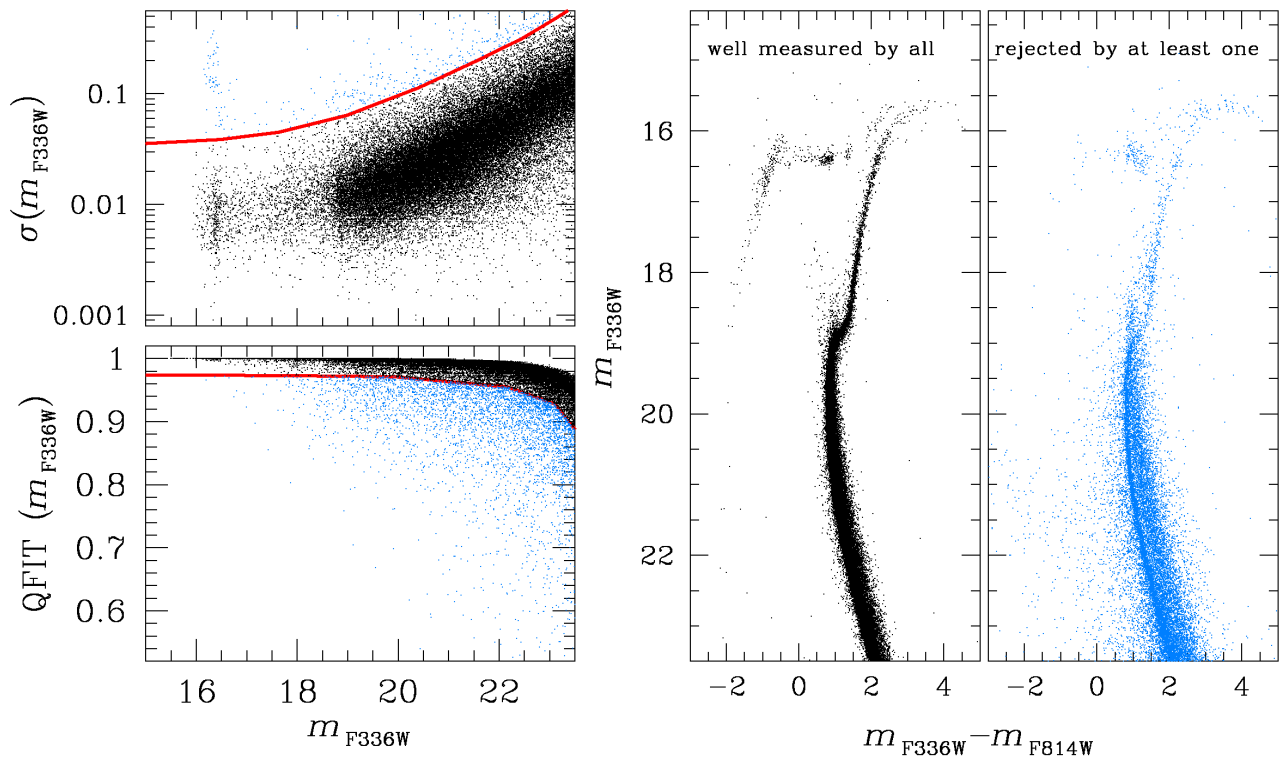}
\caption{Selection of well measured stars based on photometric errors
  (top-left panel) and \texttt{QFIT} (bottom-left panel), for the
  filter F336W: in black the stars that satisfy the selection
  criteria, in azure the rejected stars. Similar selections have been
  performed for the other filters. The CMD in the central panel shows the
  stars that passed the selections in the F336W and F814W filters; the
  right panel shows the CMD of the stars rejected by at least one of
  selection cuts.\label{fig2}}
\end{figure*}

In this work we did not use the catalogues of
\citetalias{2015AJ....149...91P}, but we reduced all the useful {\it
  HST} data of M\,15 available in the archive with new tools. These
tools allowed us to obtain lower photometric errors for faint stars
and produce artificial star catalogues, fundamental for our data
analysis.

We reduced data collected with both cameras, the {\it HST} Wide Field Channel (WFC), which
is part of the Advanced Camera for Surveys (ACS) and the UVIS
imager of the Wide Field Camera 3 (WFC3).  The observations cover the
central region of the GC M\,15. The total field of view is shown in
Fig.~\ref{fig1}, while Table~\ref{tab1} gives a detailed log of the
observations.  The analysis of the MPs hosted by M\,15 is based on the
data from GO-10775 (PI: A.\,Sarajedini), GO-12605 (PI: G.\,Piotto),
and GO-13295 (PI: S.\,Larsen).

The M\,15 catalogue belongs to the intermediate data-release of our
project. The data reduction will be presented in a forthcoming paper
(Nardiello et al. in preparation). In this analysis, we give a brief
description of the major steps of the data reduction pipeline.

  We worked on \texttt{\_flc} images, which are corrected for the
  charge-transfer inefficiency (\citealt{2010PASP..122.1035A}). For
  each image we extracted a spatial- and time-varying array of point
  spread functions (PSFs) by perturbing library
  PSFs\footnote{http://www.stsci.edu/$\sim$jayander/STDPSFs/}. We used
  these PSF arrays to extract astro-photometric catalogues from the
  images. We corrected the positions of the stars for geometric
  distortion using the routines described by
  \citet[ACS/WFC]{2006acs..rept....1A}, and by
  \citet{2009PASP..121.1419B} and
  \citet[WFC3/UVIS]{2011PASP..123..622B}.  
  We adopted the
    catalogue associated to the deepest F814W exposure ($t_{\rm
      exp}=150$\,s, \texttt{j9l954fdq}) as reference system for
    positions and we found the transformations between this master
  catalogue and all the other single-exposure catalogues, using
  six-parameter linear transformations. For each filter, the
  photometric zero-point of each filter in each individual catalogue
  is tailored to that of the deepest exposure. For each filter, we
  obtained a final catalogue containing the 3$\sigma$-clipped average
  stellar positions and fluxes in that filter (``first-pass''
  photometry, similar to that used in
  \citetalias{2015AJ....149...91P}).
  
  We extracted the ``second-pass'' photometry using the
  \texttt{FORTRAN} routine \texttt{kitchen\_sync2} (KS2, J.~Anderson
  in preparation;
  \citealt{2016ApJS..222...11S,2017ApJ...842....6B}). Using the
  images, the PSF arrays and the transformations obtained during the
  ``first-pass'' photometry, KS2 analysed all the images
  simultaneously to find and measure each source (after subtracting
  the neighbour stars) through 8 different iterations.  The KS2 software
  generated astrometric and photometric catalogues of stars using
  three different methods. A detailed description of the three methods
  is given in \citet{2017ApJ...842....6B}. In this analysis, we used only the
  method-1 measurements, which give the best results for the bright stars. The
  final catalogue of stars contains the positions $(X,Y)$, the
  magnitudes in 7 filters, and some quality parameters, such as the
  rms-based photometric errors, the quality-of-fit (\texttt{QFIT}), the number
  of images in which a star is found and the number of good
  measurements used to measure the stellar flux.  We calibrated the
  photometry into the Vega-mag system by comparing our PSF-based photometry
  against aperture photometry on \texttt{\_drc} images (which are normalised to an exposure time of
  1\,s, see \citealt{2017ApJ...842....6B} for details).  

\begin{figure*}
\includegraphics[width=0.95\textwidth]{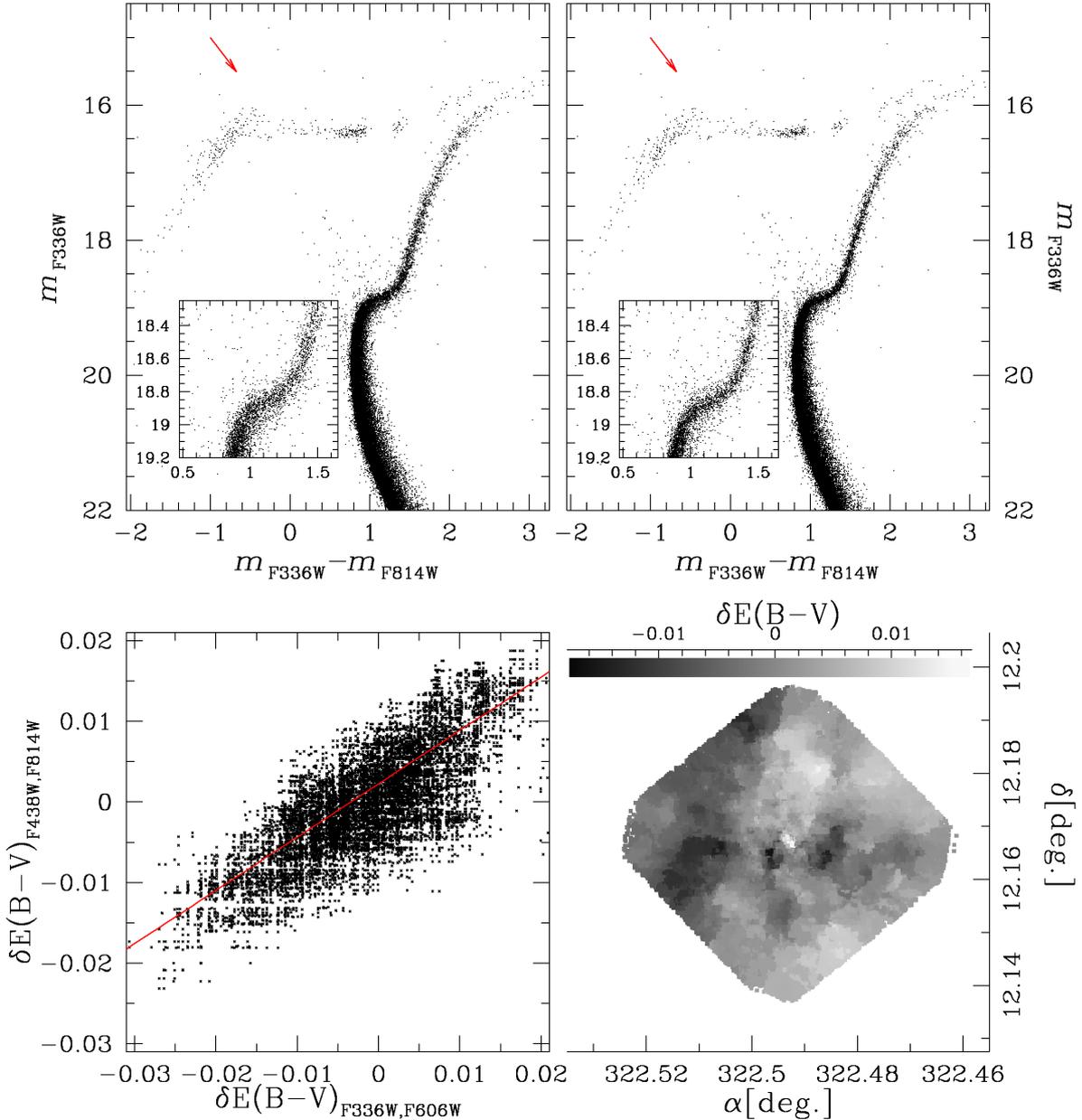}
\caption{Top panels show the $m_{\rm F336W}$ versus $m_{\rm
    F336W}-m_{\rm F814W}$ CMDs before (left panel) and after (right
  panel) the differential reddening correction; the red arrow is the
  reddening vector. The insets show a zoom-in around the SGB
  region. Bottom left panel shows a comparison between the values of
  differential reddening inferred from $m_{\rm F606W}$ versus $m_{\rm
    F336W}-m_{\rm F606W}$ and $m_{\rm F814W}$ versus $m_{\rm
    F438W}-m_{\rm F814W}$ CMDs. The red line is obtained from a
  least-squares fit of the $\delta {\rm E(B-V)}_{\rm F336W,F606W}$
  and $\delta {\rm E(B-V)}_{\rm F438W,F814W}$. Bottom right panel presents
  the map of the differential reddening. \label{fig3}}
\end{figure*}

To characterise MPs along the sequences on the CMDs of M\,15, we used
only well measured stars, selected as in \citet{2012A&A...540A..16M},
using different diagnostics such as photometric rms, \texttt{QFIT},
number of images in which the star is measured, etc. An example of
selection of well-measured stars is shown in Fig.~\ref{fig2}. 
  Left panels of Fig.~\ref{fig2} show the selections based on the
  photometric rms (top panel) and on the \texttt{QFIT} (bottom panel)
  for the filter F336W. We performed similar selections for the other
  filters. The stars that passes the selection criteria in F336W and
  F814W filters are shown in the $m_{\rm F336W}$ versus $m_{\rm
    F336W}-m_{\rm F814W}$ CMD of the central panel of Fig.~\ref{fig2};
  right panel shows the stars that are rejected by at least in one
  filter.

We corrected the magnitudes for differential reddening (DR) using the
procedure described in \citet{2012ApJ...744...58M}. Briefly, in a
given CMD (where MPs are not evident), for each star in our catalogue,
we selected the closest 50 cluster stars and measured their colour
offset from the fiducial cluster sequence (along the reddening direction).
The average and the standard deviation of the offsets was
assumed to be the local estimate of the DR $(\delta{\rm E(B-V)})$ and
the error in this estimate is considered to be the correction error for the target star. To better constrain the
variation of reddening in our field of view, we used the technique
described in \citet{2015MNRAS.446.1672M}, based on the comparison of
$\delta{\rm E(B-V)}$ (i.e. the variation of the reddening from the
average value of ${\rm E(B-V)}=0.1$) obtained from two different CMDs.
The bottom left panel of Fig.~\ref{fig3} shows the procedure: we compared
$\delta{\rm E(B-V)}$ obtained from $m_{\rm F606W}$ versus $m_{\rm
  F336W}-m_{\rm F606W}$ and $m_{\rm F814W}$ versus $m_{\rm
  F438W}-m_{\rm F814W}$ CMDs, and then performed a least-squares fit to
obtain the final value of $\delta{\rm E(B-V)}$ (red line).  The right panel
illustrates the variation of $\delta{\rm E(B-V)}$ over our field of
view.  The top panels show the $m_{\rm F336W}$ versus $m_{\rm F336W}-m_{\rm
  F814W}$ CMDs before (left panel) and after (right panel) the
correction. The correction is more evident at the SGB level, because
that is where the sequence is almost orthogonal to the reddening vector.

\subsection{Artificial-star tests}
\label{sec:ass}

In this analysis, we used artificial stars (ASs) for many different purposes: to
determine the completeness level of the analysed stars, to estimate
the impact of blends in the observed data, and to measure the fraction
of stars belonging to each population hosted by M\,15.

We produced ASs only for RGB stars, which are our main targets here.  We covered a
range of magnitude between $m_{\rm F814W}=14.20$ and $m_{\rm
  F814W}=17.30$. We generated 200\,000 ASs with a flat luminosity
function in F814W and with colours that lie along the RGB fiducial
lines in the $m_{\rm F814W}$ versus $m_{\rm X}-m_{\rm F814W}$ CMDs,
where X represents one of the available filters. The ASs have a Gaussian
spatial distribution, centred on M\,15 and with $\sigma
= 70$\, arcsec.
The software added one AS at a time to each image with the appropriate
position and flux, and then searched for the star and measured it using the same procedures
adopted for real stars and giving the same outputs.

We studied the level of completeness at different magnitudes and
radial distances from the centre of the cluster.  We considered an
artificial star to be recovered if the difference between the input and
output positions is less than 0.5 pixel and if the difference between
the input and output $m_{\rm F814W}$ magnitude is less than 0.75
mag. We found that the completeness for RGB stars is between 93\% (in
the central region) and $99$\% ($\gtrsim 1$\,arcmin from the center
of M\,15).

\begin{figure}
\includegraphics[bb=62 403 333 655, width=0.45\textwidth]{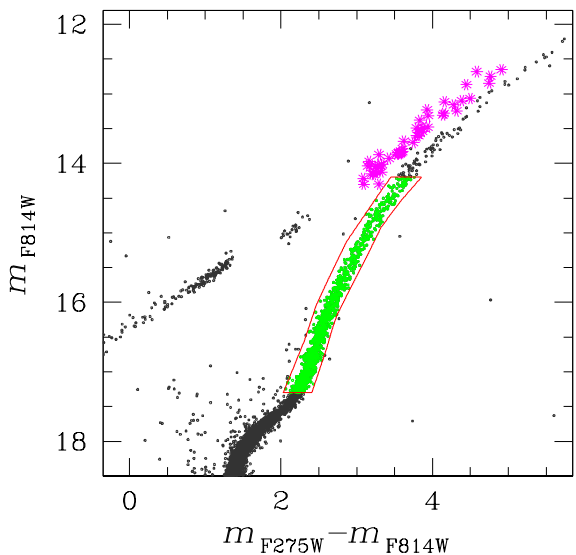}
\caption{Selection of RGB (green) and AGB  (magenta) stars based on the $m_{\rm F814W}$ versus
  $m_{\rm F275W}-m_{\rm F814W}$ CMD. \label{fig4}}
\end{figure}

\begin{figure*}
\includegraphics[width=\textwidth]{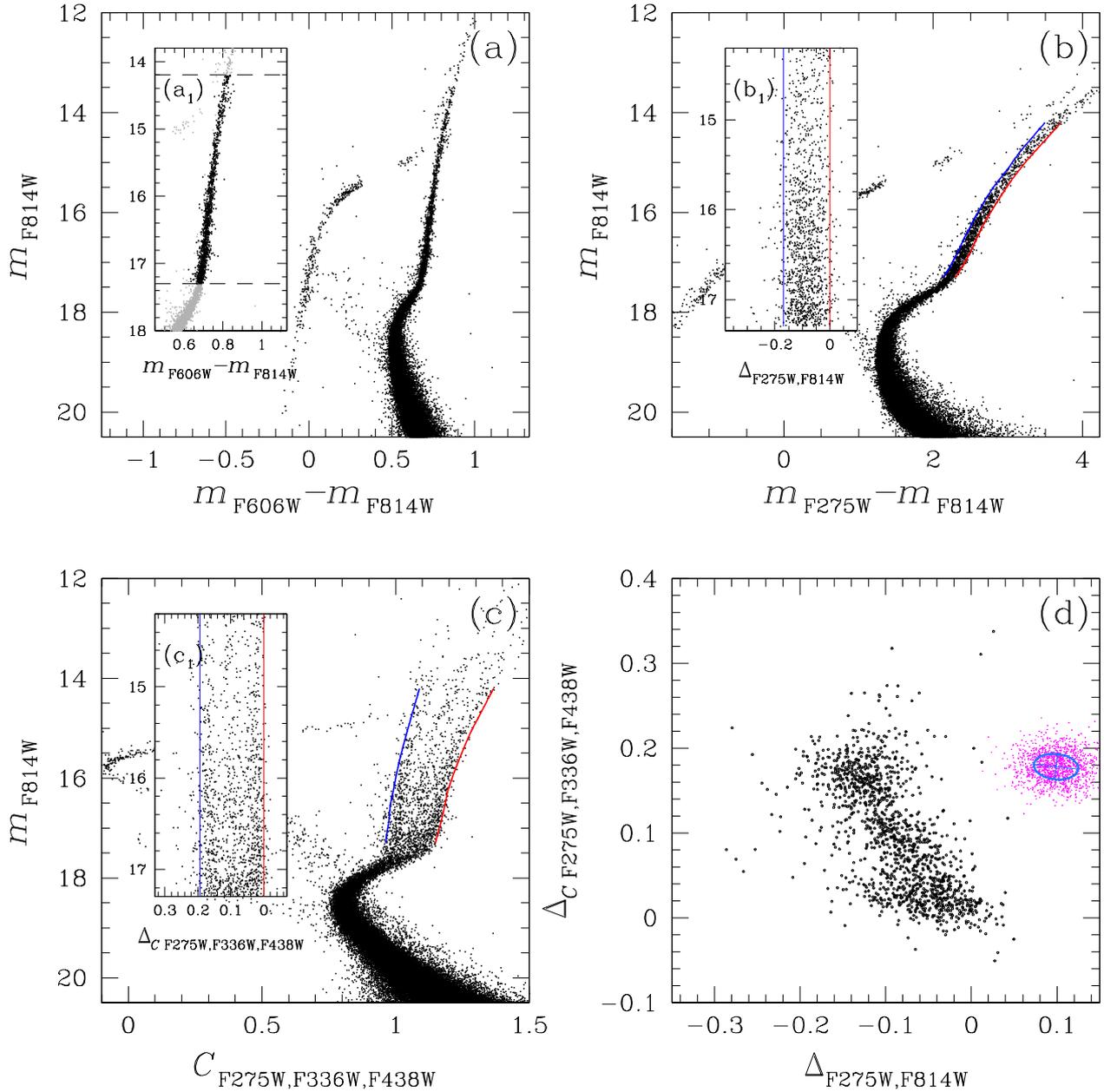}
\caption{This figure illustrates the procedure adopted to derive the
  $\Delta_{C {\rm F275W,F336W,F438W}}$ versus $\Delta_{\rm
    F275W,F814W}$ ChM of RGB stars in M\,15. {\it Panel (a)}: the
  $m_{\rm F814W}$ versus $m_{\rm F606W}-m_{\rm F814W}$ CMD of
  M\,15. The inset panel (a$_1$) shows a zoom-in on the RGB region:
  grey and black points represent all the well-measured stars and the
  selected RGB stars, respectively. {\it Panels (b) and (c)}: the
  $m_{\rm F814W}$ versus $m_{\rm F275W}-m_{\rm F814W}$ and the $m_{\rm
    F814W}$ versus $C_{\rm F275W, F336W, F438W}$ CMDs of M\,15,
  respectively. The blue and red lines are the fiducials used to
  verticalize the RGB (inset panels (b$_1$) and (c$_1$)), and
  correspond to the 4th and 96$^{\rm th}$ percentile of the RGB colour
  distribution. {\it Panel (d)}: $\Delta_{C {\rm F275W,F336W,F438W}}$
  versus $\Delta_{\rm F275W,F814W}$ ChM, obtained as explained in the
  text. The points in magenta are the distribution of the
  observational errors; the azure ellipse include the 68.27 per cent
  of the points.  \label{fig5}}
\end{figure*}

\begin{figure*}
\includegraphics[width=0.95\textwidth]{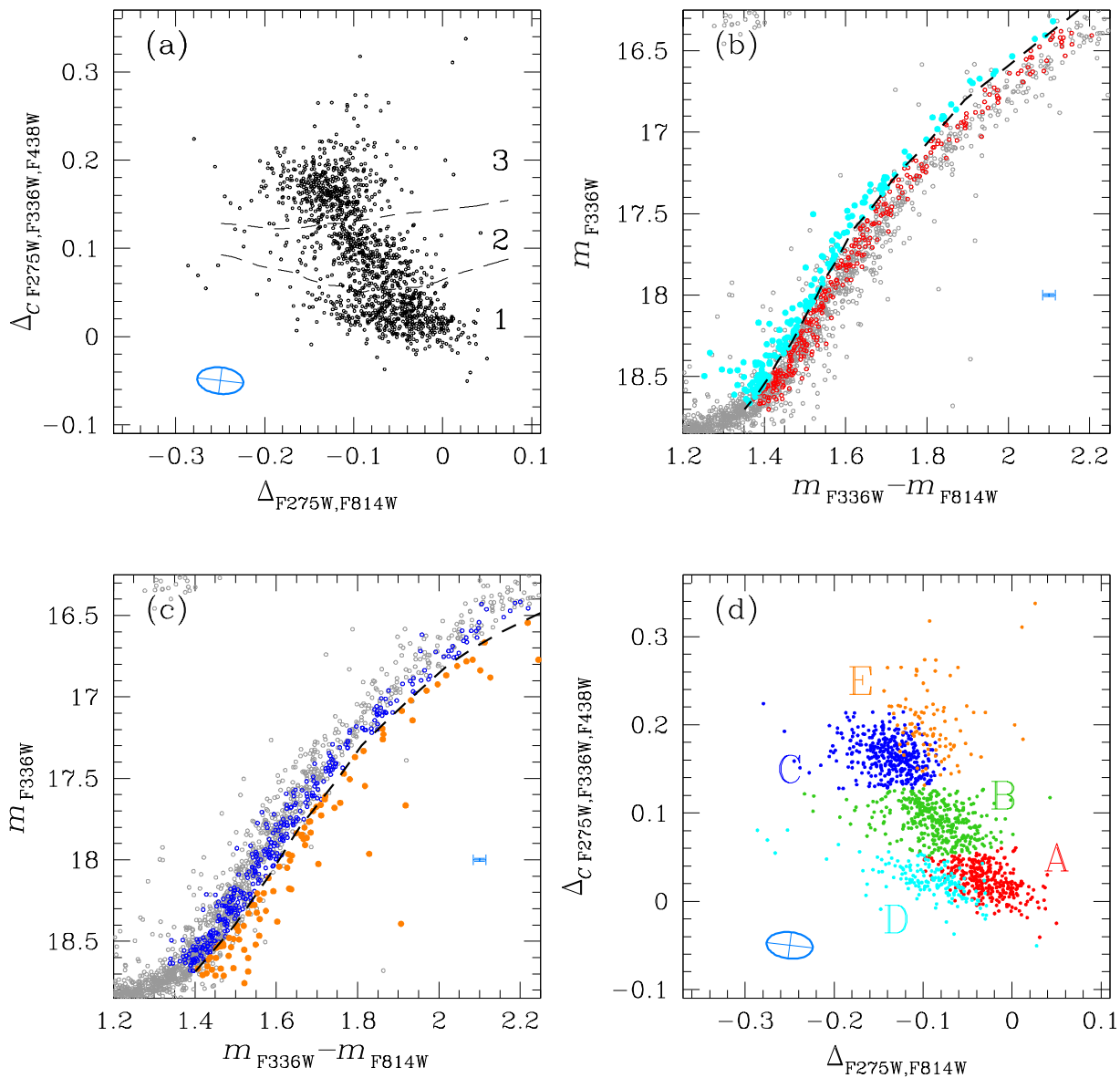}
\caption{Procedure adopted to identify the different populations
  hosted by M\,15.
  The dashed lines overimposed on the ChM shown in the panel (a)
    are used to identify the three groups of RGB-1, RGB-2 and RGB-3
    stars. Panels (b) and (c) show the $m_{\rm F336W}$ vs.\,$m_{\rm
      F336W}-m_{\rm F814W}$ CMD of RGB stars. The dashed lines plotted
    in the panels (b) and (c) are used to separate the two
    sub-populations A and D of the RGB-1 and the sub-populations C and
    E of the RGB-3, respectively. In panel (d) we use red, green,
    blue, cyan, and orange colours to represent stars in the
    populations A, B, C, D, and E, respectively, in the ChM. See text
    for details.
   \label{fig6}}
\end{figure*}

\begin{figure*}
\includegraphics[width=\textwidth, bb=17 361 589 710, width=0.9999\textwidth ]{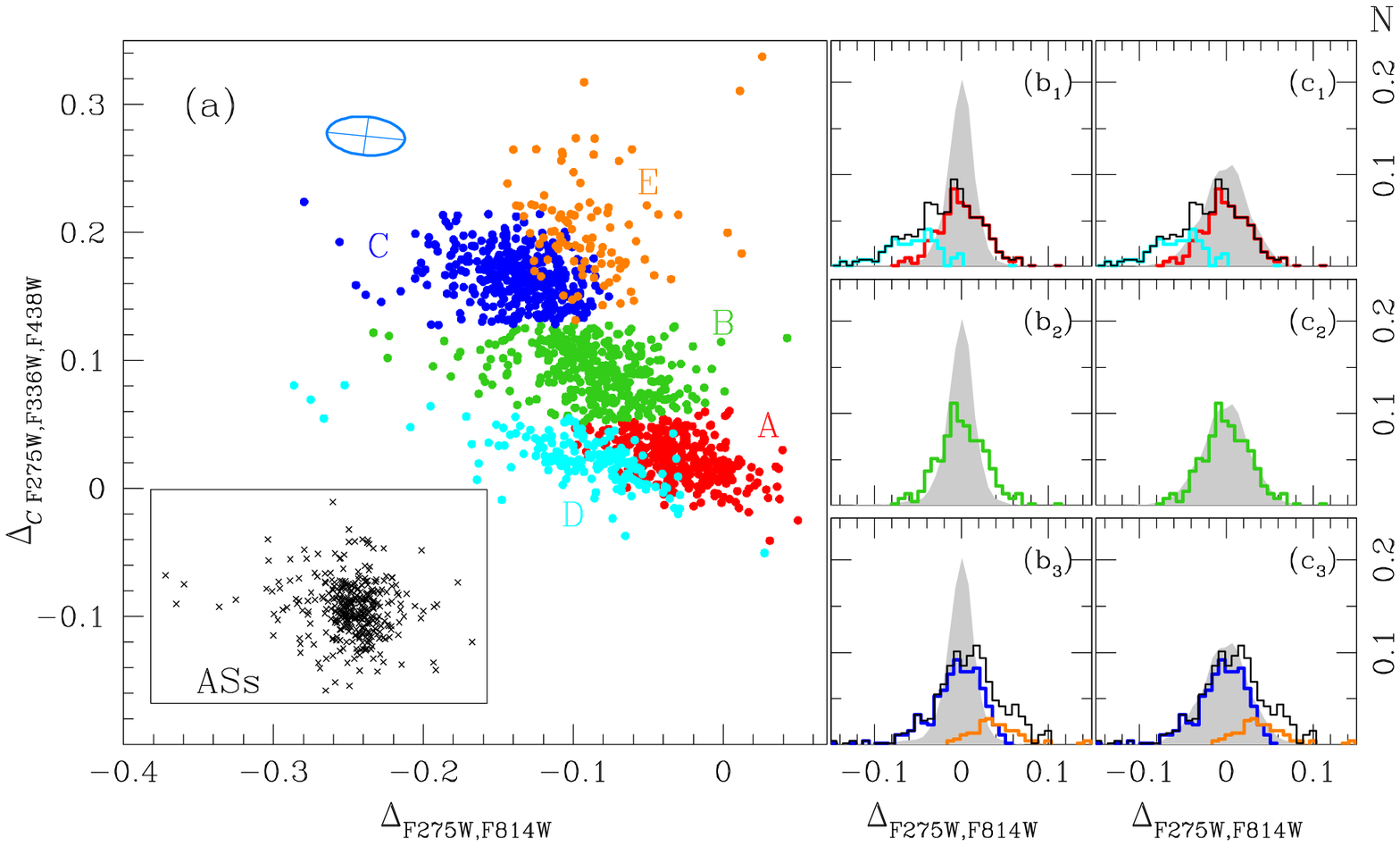}
\caption{{\it Left-panel:} Observed and simulated (in black) ChMs;
  colour codes for each observed population are as in Fig.~\ref{fig6}..
  The azure ellipse is indicative of the observational errors. {\it
    Right-panels:} Comparison between the $\Delta_{\rm F275W,F814W}$
  colour distributions of ASs (grey) and POP\,A, POP\,D, and
  POP\,A+POP\,D (in red, cyan, and black, respectively, panels (b1)
  and (c1)), POP\,B (green, panels (b2) and (c2)), and POP\,C, POP\,E,
  and POP\,C$+$E (in blue, orange, and black, respectively, panels
  (b3) and (c3)).  Panels (b) show the colour distribution of ASs as
  measured by the software KS2; in panels (c) we broadened the ASs
  colour distribution to take into account of the underestimation of
  AS photometric errors.  All the distributions are normalised to the
  total number of considered stars. See text for
  details.  \label{fig7}}
\end{figure*}

\section{The chromosome maps and the multiple stellar populations in M\,15}
\label{sec:chrm}

In \citet[hereafter
  \citetalias{2015MNRAS.447..927M}]{2015MNRAS.447..927M} and
\citetalias{2017MNRAS.464.3636M}, we introduced a pseudo 
two-colour
diagram, or chromosome map (ChM), in order to separate stars with
1G- and 2G-type abundance patterns.  We used this new tool to identify
1G and 2G stars in 57 GCs.  In this work, we exploit the ChM to
identify MPs along the RGB of M\,15.  We then calculated the
number of stars in each population and studied their radial distribution.

In the following we describe the method we used to construct the ChMs by using
the $m_{\rm F275W}$, $m_{\rm F336W}$, $m_{\rm F438W}$, and $m_{\rm F814W}$
filters.
For a detailed description we refer to
\citetalias{2017MNRAS.464.3636M}.

 We started by identifying isolated RGB stars using the $m_{\rm F814W}$ versus $m_{\rm
   F275W}-m_{\rm F814W}$ CMD (see Fig.~\ref{fig4}); 
 we excluded stars that are more distant than 0.17 mag in colour from
 the fiducial line of the RGB. The objects that we have rejected
 include evolved blue stragglers, photometric blends, and binaries
 with mass ratio close to one. The selected sample includes 1,309 RGB
 stars, which are marked with green points in Fig.~\ref{fig4}.
 
The procedure adopted to derive the $\Delta_{C {\rm
    F275W,F336W,F438W}}$ versus $\Delta_{\rm F275W,F814W}$ ChM is
 illustrated in Fig.~\ref{fig5}.  In panel (a$_1$) of Fig.~\ref{fig5}
we highlight in black the RGB stars. We divided the RGB
into a set of F814W magnitude bins (of width $\delta m =0.4$\,mag). We
divided each bin in $N$ sub-bins of width $\delta m/3$.

We calculated the 4$^{\rm th}$ and the 96$^{\rm th}$ percentiles of the
colour distribution and the mean F814W magnitude in each interval
$m_{\rm F814W}^i<m_{\rm F814W}<m_{\rm F814W}^i+\delta m$, with
$i=1,...,N$. We used a 3-point boxcar to smooth the 4$^{\rm th}$ and the 96$^{\rm th}$ points and then
interpolated them with a spline for obtaining the blue (4$^{\rm th}$
percentile) and red (96$^{\rm th}$ percentile) lines shown in panels
(b) and (c), respectively, for the cases of 
$m_{\rm F275W}-m_{\rm F814W}$ colour and 
 $C_{\rm F275W,F336W,F43W}=(m_{\rm F275W}-m_{\rm F336W})
   -(m_{\rm F336W}-m_{\rm F438W})$ pseudo-colour, from Fig.~\ref{fig5}.

For each colour $C$, we computed the observed RGB width $W^{\rm
  obs}_C$ as the difference between the colours of red and the blue
lines at 2.0 F814W magnitudes above the MS turn-off ($m_{\rm F814W,
  TO} = 18.875 \pm 0.008$). We subtracted in quadrature the photometric
and DR-correction errors to $W^{\rm obs}_C$, obtaining the intrinsic RGB width
$W$.

For each star we computed the quantities:
\begin{equation}
  \Delta_{\rm F275W,F814W} = W_X\frac{X-X_{\rm fiducialR}}{X_{\rm fiducialR}-X_{\rm fiducialB}}
\end{equation}

and 

\begin{equation}
  \Delta_{C {\rm F275W,F336W,F438W}} = W_Y\frac{Y_{\rm fiducialR}-Y}{Y_{\rm fiducialR}-Y_{\rm fiducialB}}
\end{equation}
where $X=(m_{\rm F275W}-m_{\rm F814W})$, $Y=C_{\rm F275W, F336W,
  F438W}$, ``fiducialR'' and ``fiducialB'' are the red and blue
fiducial lines of panels (b) and (c) of Fig.~\ref{fig5}. The
verticalized RGBs are shown in the inset panels (b$_1$) and
(c$_1$). Panel (d) shows the ChM $\Delta_{C {\rm F275W,F336W,F438W}}$
versus $ \Delta_{\rm F275W,F814W}$ colours; the points in magenta are
the distribution of the observational errors in the ChM, which are a
combination of photometric and DR errors.

\subsection{Multiple stellar populations along the RGB}
\label{sub:mpops}

The most prominent features of the ChM plotted in
  Fig.~\ref{fig5} are three groups of stars clustered around
  $\Delta_{C {\rm F275W,F336W,F438W}} \sim$ 0.03, 0.10, and 0.18.  In
   panel (a) of Fig.~\ref{fig6} we have drawn by hand two dashed
  lines to select the corresponding main RGBs of M\,15, which we call
  RGB-1, RGB-2, and RGB-3.
  
Noticeably, the triple RGB of M\,15 was clearly visible in the $m_{\rm
  F336W}$ vs.\,$C_{\rm F275W,F336W,F438W}$ pseudo CMD shown in the
\citetalias{2015AJ....149...91P} (see their Fig.~22) and was studied
by \citet{2015ApJ...804...71L} by using F343N and F555W WFC3/UVIS
photometry. In the following, we demonstrate that the MP phenomenon in
M\,15 is even more complex than previously believed.

In \citetalias{2017MNRAS.464.3636M} we discovered that the ChMs of all
the analysed GCs, including M\,15, host two distinct groups of 1G and
2G stars. The first generation of M\,15 identified in \citetalias{2017MNRAS.464.3636M}
corresponds to the RGB-1, while the second generation includes the
sub-populations of RGB-2 and RGB-3 stars.

The ChM of M\,15 reveals  a complex morphology which is not adequately reproduced by only three simple stellar populations.
 As discussed in \citetalias{2017MNRAS.464.3636M},
the $\Delta_{\rm F275W,F814W}$ broadening of 1G stars is much larger
than what we would expect from observational errors alone, thus indicating
that the RGB-1 is not consistent with being a single, simple population.  Moreover,
we note a poorly-populated group of RGB-3 stars spread around
$\Delta_{\rm F275W,F814W} \sim -0.1$ and $\Delta_{C {\rm
    F275W,F336W,F438W}} \sim 0.25$.

To investigate the morphology of the RGB-1 we show in 
panel (b) of Fig.~\ref{fig6} a zoom of the $m_{\rm F336W}$
vs.\,$m_{\rm F336W}-m_{\rm F814W}$ CMD around the RGB.
In this CMD, as well, the colour spread of RGB-1 stars, which are
represented with coloured circles, is much larger than what we expect
from observational errors alone ($\sim$0.015 mag for these bright stars),
thus confirming that the RGB-1 is not composed by a simple homogeneous population.   We drew by hand the black dashed line to separate
the two sub-populations A and D of RGB-1 stars and colored them red and
cyan, respectively.

Similarly, in panel (c) of Fig.~\ref{fig6} we selected two
sub-populations, C (blue circles) and E (orange circles), of RGB-3
stars.  The location on the ChM of stars in the five populations of
M\,15 is provided in panel (d) of Fig.~\ref{fig6}, where we indicate
RGB-2 stars as population B.  In the following we provide further
evidence that the stellar groups A+D and C+E cannot be considered as simple stellar populations.

\begin{figure*}
\begin{center}
\includegraphics[bb=43 326 422 522, width=0.9\textwidth]{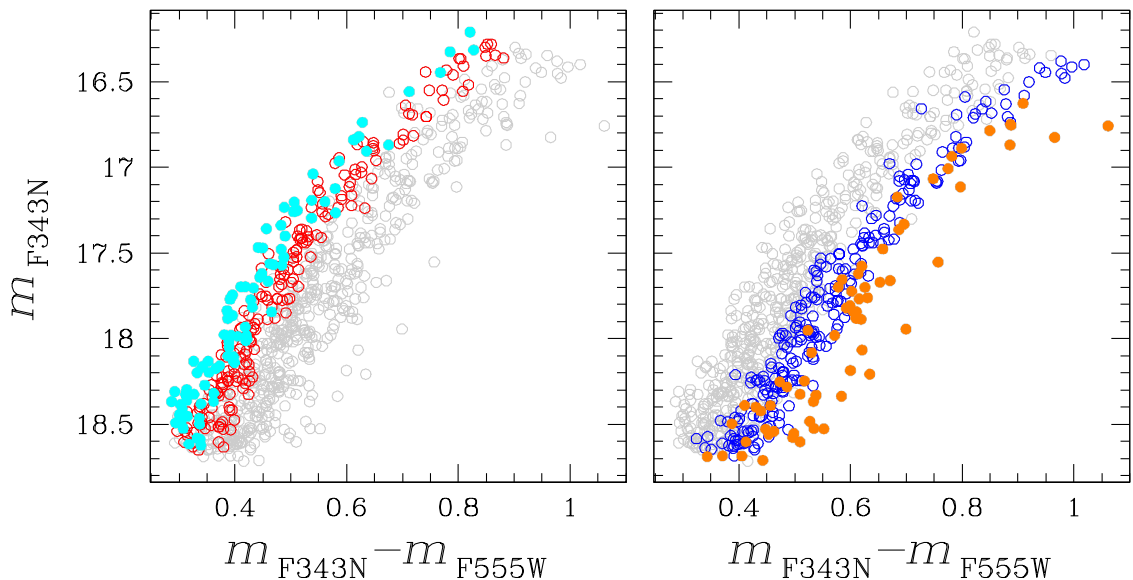}
\caption{The $m_{\rm F343N}$ vs.\,$m_{\rm F343N}-m_{\rm F555W}$ CMDs of
  RGB stars in M\,15. Population A and D stars are coloured red and
  cyan. Blue and orange colours indicate populations C and E,
  respectively.
   \label{fig8}}
\end{center}
\end{figure*}

\begin{figure*}
\includegraphics[width=0.75\textwidth]{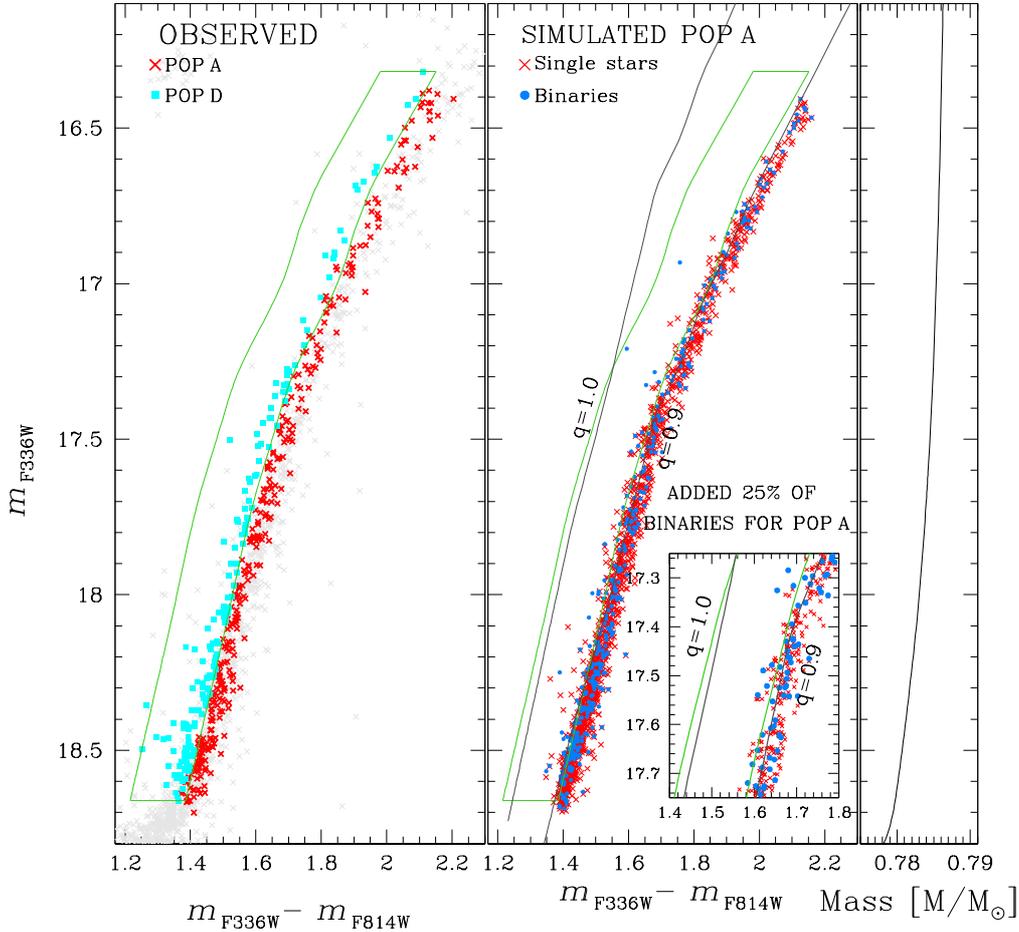}
\caption{Procedure used to simulate the RGB binary population. {\it
    Left panel:} observed RGB sequences for POP\,A (red crosses) and
  POP\,D (cyan squares). The green region includes  most of POP\,D
  stars. {\it Middle panel: } simulated RGB sequence for POP\,A (red
  crosses) and simulated POP\,A binary sequence (azure circles). Black lines
  are the locii of binary stars having mass ratios $q=0.9$ and
  $q=1.0$. In this simulation we added a fraction of stars equal to
  the 25\% of the total number of simulated RGB stars. {\it Right
    panel: } stellar mass as a function of the F336W
  magnitude. \label{fig9}}
\end{figure*}

\begin{figure*}
\includegraphics[bb=26 320 432 700,width=0.75\textwidth]{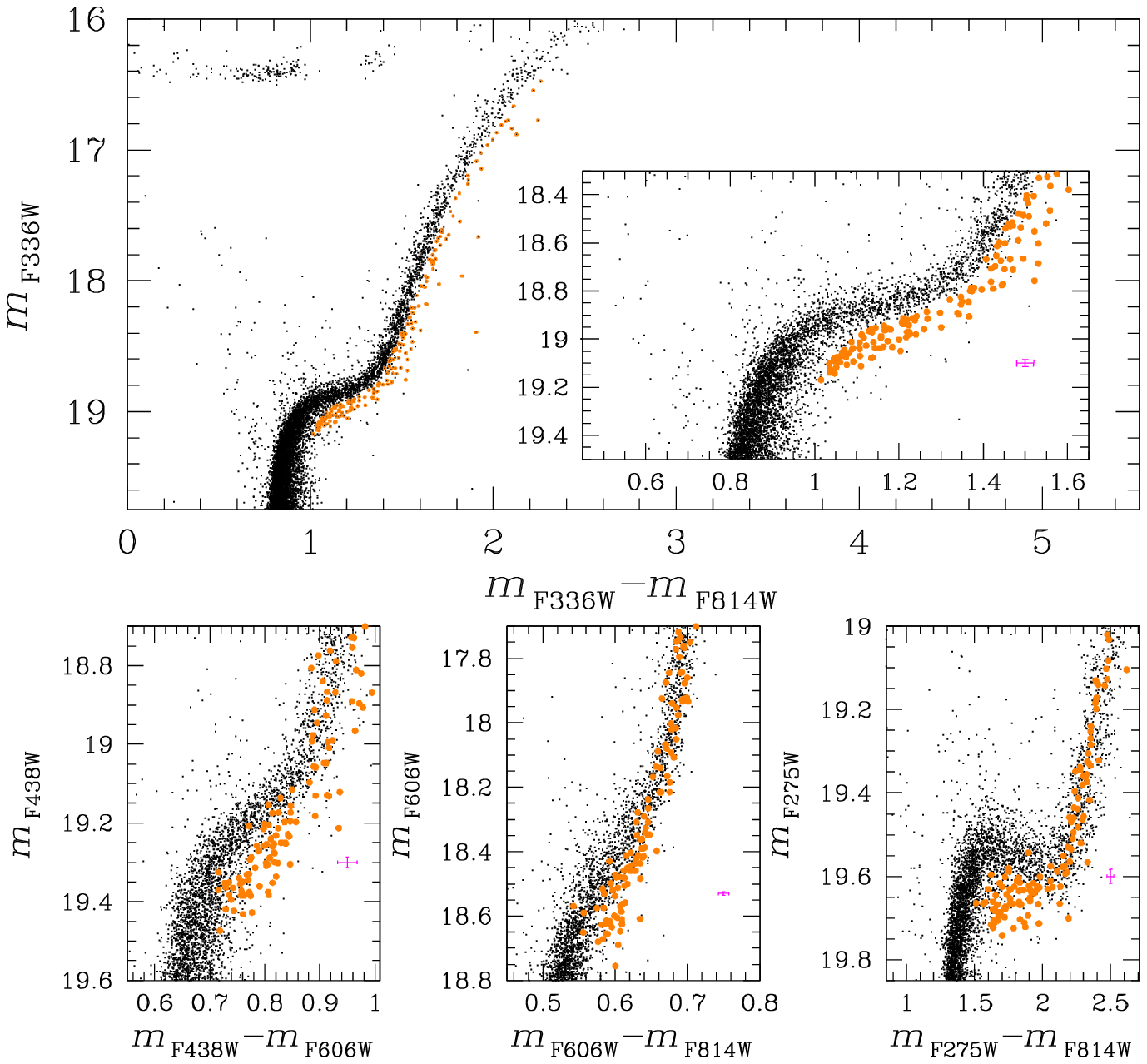}
\caption{{\it Top panel: }  $m_{\rm F336W}$ versus $m_{\rm
    F336W}-m_{\rm 814W}$ CMD of M\,15, with the POP\,E coloured in
  orange. The inset shows a zoom-in around the SGB region. The faint
  SGB, possibly associated with POP\,E, is highlighted in
  orange. {\it Bottom panels: } $m_{\rm F438W}$ versus $m_{\rm
    F438W}-m_{\rm 606W}$ (left), $m_{\rm F606W}$ versus $m_{\rm
    F606W}-m_{\rm 814W}$ (middle), and $m_{\rm F275W}$ versus
  $m_{\rm F275W}-m_{\rm 814W}$ (right) CMDs around the SGB
  region. The faint SGB stars selected in the top panel are coloured in
  orange in these CMDs as well. The magenta crosses show the 
  photometric errors at the SGB level. \label{fig10}}
\end{figure*}

\subsection{Populations D and E}
\label{ssec:blends}
In this Section  we show how the morphology of the ChM of
  M15 is better reproduced by adding the two stellar populations
  (POPs) D and E.

We first show that POP\,D and POP\,E stars are neither photometric
blends nor stars with large observational errors.  To do this, we
adopted the procedure illustrated in Fig.~\ref{fig7}, which is based
on the comparison between the observed ChM of M\,15 and the 
artificial-star simulated
ChM of a single stellar population. To derive the latter, we added to
the $\Delta_{\rm F275W,F814W}$ and $\Delta_{C {\rm
    F275W,F336W,F438W}}$ quantities derived from ASs the errors
associated with the differential-reddening corrections.  The observed
and the simulated ChMs are plotted in the panel (a) of
Fig.~\ref{fig7}.
   
In panels (b1) to (b3), we show that the $\Delta_{\rm F275W,F814W}$
histogram distributions for populations A$+$D, B, and C$+$E are
broader than the corresponding simulated distributions. A possible
explanation is that each population A$+$D, B, and C$+$E host stars
that are not chemically homogeneous.  As an alternative, the
difference between the dispersion of the simulated and observed
$\Delta_{\rm F275W,F814W}$ distributions could indicate that the
magnitude errors inferred from ASs are simply lower limits on the real
uncertainties on magnitude measurements.

To demonstrate that the groups of A$+$D, and C$+$E stars are not
simple populations, we first assumed that the magnitude errors
inferred from ASs are underestimated.  To compensate for this fact, we
broadened the distribution of ASs by adding to each AS a random noise
in colour in such a way that the simulated histogram distribution
matches the corresponding distribution of POP\,B stars.  In doing this,
we assume that POP\,B stars are chemically homogeneous, which might
not be true. If the hypothesis is false, we are simply overestimating
the photometric errors making the following conclusion on the reality
on the A, D, C, and E populations even stronger.

Results are illustrated in panels (c1) to (c3) of Fig.~\ref{fig7} and
show that the histograms of populations A$+$D and C$+$E (in black)
exhibit a blue and a red tail, respectively.  These tails are not present in
the simulated histograms. Moreover, the AS dispersion is consistent
with the observed dispersion of POP\,A and POP\,C only. We conclude
that the stellar groups A--E selected in Fig.~\ref{fig6} are not
artifacts but correspond to distinct stellar populations.

As an alternative method   to demonstrate that the observed spreads in the ChM
  in the region corresponding to populations A and C are real, in
Fig.~\ref{fig8} we analyse their position in the $m_{\rm F343N}$
vs.\,$m_{\rm F343N}-m_{\rm F555W}$ CMD.  The five sub-populations of
M\,15, including populations D and E, have been identified by using
photometry in the F275W, F336W, F438W, and F814W bands (see
Section~\ref{sub:mpops}) whereas the CMD of Fig.~\ref{fig8} comes from
a dataset, which is independent from that used in
Section~\ref{sub:mpops}.

If the POP\,A$+$POP\,D are in fact a single population and the colour
spread of the RGB made of POP\,A$+$POP\,D stars is due to
photometric errors alone, then a star that is red (or blue) relative
to the sequence in the diagrams
of Section~\ref{sub:mpops} should have the same probability of being
either red or blue in the $m_{\rm F343N}$ vs.\,$m_{\rm F343N}-m_{\rm
  F555W}$ CMD. By contrast, the fact that the two populations form two
distinct sequences in the $m_{\rm F343N}$ vs.\,$m_{\rm F343N}-m_{\rm
  F555W}$ CMD, demonstrates that the colour spread of the
POP\,A+POP\,D RGB is intrinsic and that the POP\,D stars have
different photometric properties with respect to POP\,A
(\citealt{2009ApJ...697L..58A,2010ApJ...709.1183M,2015A&A...573A..70N}). Similar
arguments demonstrate that POP\,C and POP\,E are truly distinct sub-populations of
M\,15.

 In this work we assume that POP\,A and POP\,D are two discrete
  populations, each one characterized by specific chemical
  properties. Because photometric errors do not allow us to totally
  split the colours of POP\,A and POP\,D in all colour-magnitude
  and two-colour diagrams, the hypothesis that the spread along the
  $\Delta_{\rm F275W,F814W}$ colour in the ChM is due to a continuous
  (and not discrete) variation of chemical elements among the stars of
  the two populations can not be excluded.  Bearing in mind the above consideration, in the following analysis we will consider the scenario in which M15 hosts 5 distinct stellar populations.

 Finally, we excluded that the colour distribution of
  POP\,A+POP\,D observed in the ChM is due to temperature
  effects. From the models we expect that stars located at the basis
  and at the top of the RGB sequence adopted to obtain the ChM have a
  difference in temperature of $\delta T_{\rm eff}\sim 1000$\,K. A
  colour dependence by $\delta T_{\rm eff}$ would result in a dependence of the
  ChM on the luminosity, that is not observed.

\begin{figure*}
\includegraphics[width=0.90\textwidth, bb=20 242 581 716]{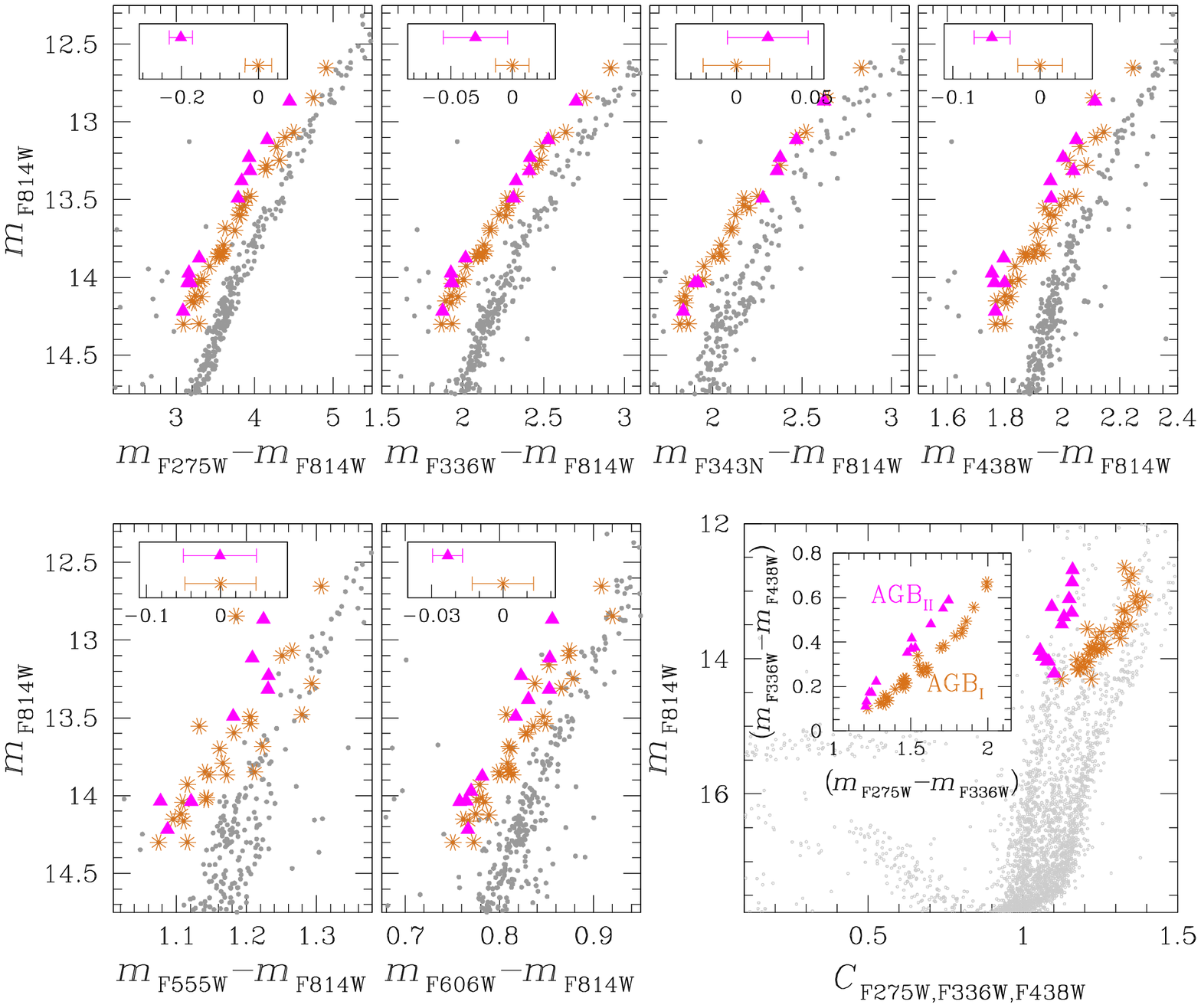}
\caption{Analysis of AGB stars. Top-panels and first two bottom panels
  show the $m_{\rm F814W}$ versus $m_{\rm X}-m_{\rm F814W}$ CMDs, with
  X=F275W, F336W, F343N, F438W, F555W and F606W; the inset panels show
  the average colour difference between AGB$_{\rm I}$ and AGB$_{\rm
    II}$ sequences. The right-hand bottom panel shows the $m_{\rm
    F814W}$ versus ${C_{\rm F275W,F336W,F438W}}$ pseudo-CMD for all
  the stars (grey points) and for AGB stars (coloured points). Inset
  panel is the $m_{\rm F336W}-m_{\rm F438W}$ versus $m_{\rm
    F275W}-m_{\rm F336W}$ two-colour diagram of AGB stars.  In all
  the diagrams, brown starred dots and magenta triangles represent the
  AGB$_{\rm I}$ and AGB$_{\rm II}$ stars,
  respectively.  \label{fig11}}
\end{figure*}

\subsection{The impact of binaries on the CMD}
The POP\,D stars always form a sequence that is on average bluer than the
POP\,A sequence; this fact might lead us to think that the POP\,D
stars are binaries of POP\,A. We evaluated this hypothesis by performing a
simulation of RGB POP\,A binary population using the procedure
described in \citet{2012A&A...540A..16M}. We used ASs to simulate the
$m_{\rm F336W}$ versus $m_{\rm F336W}-m_{\rm F814W}$ CMD of
POP\,A. For each star in POP\,A we considered 5 ASs with F814W
magnitudes within $\pm 0.10$ and with radial distances within 25
pixels from the target star; in this way we took into account the
contribution of possible blends to the final simulated CMD.  As
we did previously in Section~\ref{ssec:blends}, we also broadened the colour to
simulate the spread of the sequence due to photometric
errors. Figure~\ref{fig9} shows the observed (left panel) and the
simulated (middle panel) CMDs for POP\,A (red crosses).  We derived
the mass of the simulated RGB stars using the mass-luminosity relation
of \citet{2007AJ....134..376D}. We used an isochrone with age 13.25
Gyr, [Fe/H]$=-2.33$, [$\alpha$/Fe]$=0.20$ and primordial helium. The
mass-luminosity relation is shown in the right panel of
Fig.~\ref{fig9}.

We chose to add to the simulated POP\,A a fraction of binary stars
equal to 25\% of the total number of simulated POP\,A stars, with a
flat mass (and luminosity) distribution. For each of these stars,
having mass $\mathcal{M}_1$, we calculated the mass of the secondary
star as $\mathcal{M}_2=q \times \mathcal{M}_1$, with a mass ratio
$0<q\leq 1$.  For the simulation we adopted a flat distribution for the mass ratios
  $q$. Using the mass-luminosity relation illustrated in right panel
of Fig.~\ref{fig9}, we obtained the luminosity of the secondary
component. We added the fluxes of the two components in F336W and
F814W bands, transformed them in magnitudes and replaced the original
star in the CMD with this binary system. In Fig.~\ref{fig9} the binary
stars are plotted in azure. When $q\lesssim 0.95$, the secondary star
is a main sequence star and the contribution of the secondary
component to the total binary luminosity is negligible. For this
reason, most of simulated binary stars ($\sim 97\%$) are
indistinguishable from the simulated single-star RGB sequence. Only
few binary stars fall inside the region (marked in green in the left
and middle panels of Fig.~\ref{fig9}) that contains most of POP\,D
stars (cyan squares in the left panel of Fig.~\ref{fig9}).  In
conclusion, assuming a flat mass distribution for the POP\,A binary
components we expect that the bulk of RGB binaries is formed by a RGB
+ a MS star.  Therefore, the probability that all POP\,D stars are
binaries with mass ratio $>0.95$ is
low\footnote{\citet{2012A&A...540A..16M} found that MS+MS binaries
  with $0.833 < q \le 1.000$ are about 0.3\% of MS stars}, we can also
exclude the hypothesis that POP\,D represents a sequence of POP\,A
binaries. With similar reasoning, POP\,D stars are also not binaries
of POP\,B and POP\,C.

\begin{figure*}
\includegraphics[width=1\textwidth, bb=20 298 592 578]{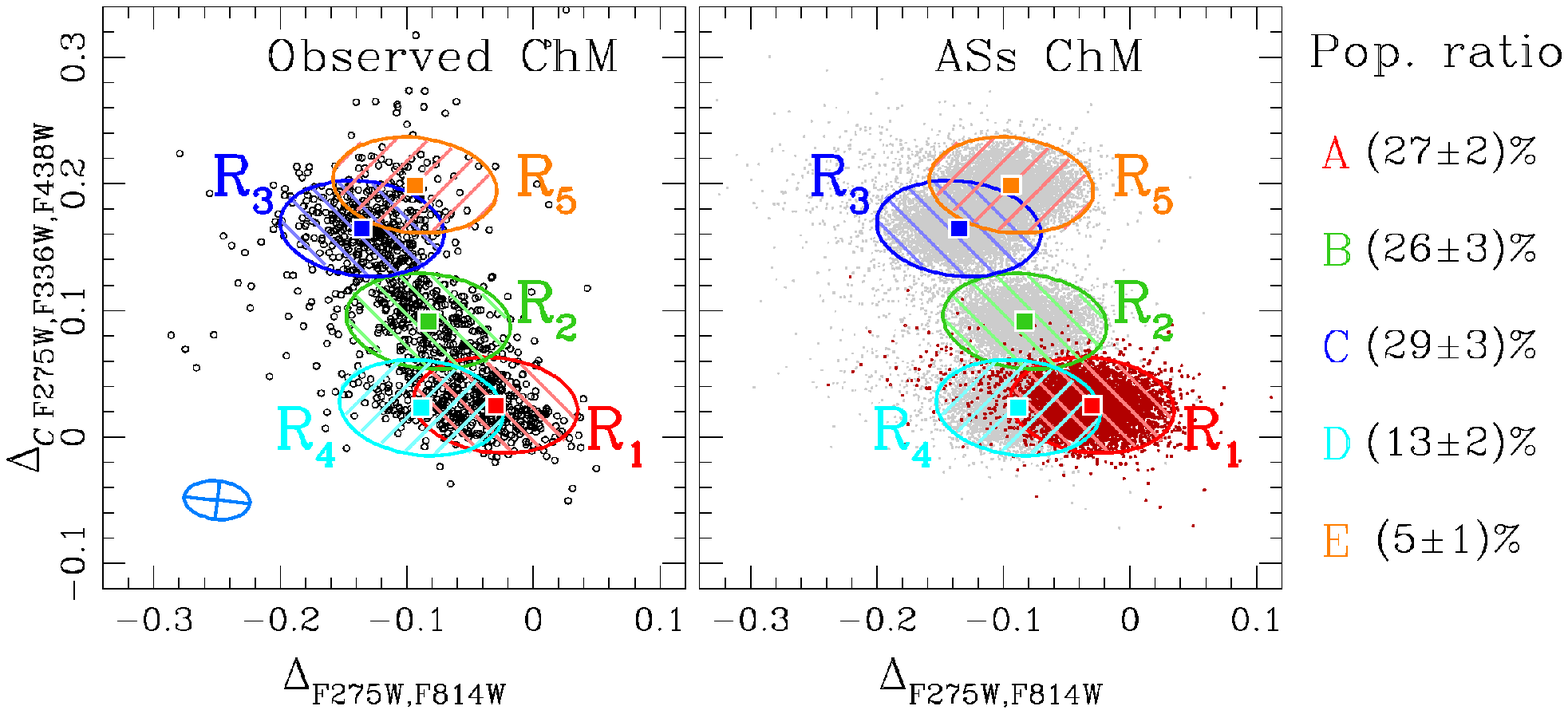}
\caption{Procedure adopted to estimate the fraction of stars in each
  population. Left panel shows the observed $\Delta_{C\,{\rm
      F275W,F336W,F438W } } $ versus $\Delta_{\rm F275W,F814W}$ ChM
  (black empty circles). The azure ellipse is indicative of the
  dispersion of a single population ($\sigma_{\rm obs}$) . The regions
  R$_i$ correspond to a size of $2.5 \times \sigma_{\rm
    obs}$. Coloured squares correspond to the centre of the identified
  populations. Right panel shows the ASs $\Delta_{C\,{\rm
      F275W,F336W,F438W } } $ versus $\Delta_{\rm F275W,F814W}$
  ChM. The regions R$_i$ have the same size and are located in the
  same positions of the regions of the left panel.  \label{fig12}}
\end{figure*}

\subsection{Multiple stellar populations along the SGB}

Figure~\ref{fig10} shows that the SGB of M\,15 is not consistent with
a single population.  Specifically, the $m_{\rm F336W}$ vs.\,$m_{\rm
  F336W}-m_{\rm F814W}$ CMD plotted in the upper panel, shows a
population of SGB stars that are spread below the bulk of the SGB
stars. These stars, which have been selected in the $m_{\rm F336W}$
versus $m_{\rm F336W}-m_{\rm F814W}$ CMD (orange dots in top-panel of 
Fig.~\ref{fig10}), define a stellar sequence that is fainter than the
majority of SGB stars in all the CMDs, including those made with
only the optical bands.  The faint SGB is clearly connected with the
POP\,E RGB stars, which exhibit redder $m_{\rm F336W}-m_{\rm F814W}$ 
colours than the remaining RGB stars with the same luminosity.

 Recently, \citet{2017ApJ...840...66G} have found a class of objects
  located in a CMD region redder than the MS and fainter than the SGB
  of many globular and open clusters. These sub-subgiant stars result
  to be evolved binary stars. The faint SGB we identify could be
  formed by sub-subgiant stars. However, we exclude this hypothesis for two
  reasons: (i) the faint SGB is connected to the POP\,E RGB; (ii) the
  number of faint SGB stars (146) is much larger than the average number of
  sub-subgiant stars observed in some GCs ($\sim 10$ stars).

\subsection{Multiple stellar populations along the asymptotic giant branch}
\label{sec:agb}
Previous papers from our group have demonstrated that the $m_{\rm F814W}$
versus $C_{\rm F275W,F336W,F438W}$ pseudo-CMD is also a powerful tool to
identify multiple populations along the AGB (e.g.\,Papers II,
III, V; \citealt{2017ApJ...843...66M}).  The bottom-right panel of
Fig.~\ref{fig11} reveals that the 45 AGB stars selected in
Fig.~\ref{fig4} distributes along two distinct sequences of AGB$_{\rm
  I}$ (brown asterisks) and AGB$_{\rm II}$ stars (magenta triangles)
in the $m_{\rm F814W}$ versus $C_{\rm F275W,F336W,F438W}$ diagram. The
selection of the two AGB groups has been performed in the $m_{\rm
  F336W}-m_{\rm F438W}$ vs.\,$m_{\rm F275W}-m_{\rm F336W}$ diagram.
The two groups of AGB$_{\rm I}$ and AGB$_{\rm II}$ contain
$(76\pm17)$\,\% and $(24\pm11)$\,\% of AGB stars, respectively.
 
Figure~\ref{fig11}, provides a collection of $m_{\rm F814W}$ versus
$m_{\rm X}-m_{\rm F814W}$ CMDs, (X$=$F275W, F336W, F343N, F438W,
F555W, and F606W) focused on the region around the AGB. 
The inset panels of the CMDs shown in Fig.~\ref{fig11} illustrate
the average colour difference between ${\rm AGB}_{\rm I}$ and ${\rm
  AGB}_{\rm II}$ sequences. ${\rm AGB}_{\rm II}$ sequence is, on
average, bluer than ${\rm AGB}_{\rm I}$ stars in all the CMDs, except
in the $m_{\rm F814W}$ versus $m_{\rm F555W}-m_{\rm F814W}$ CMD, where
the two groups share the same colour, and in the $m_{\rm F814W}$
versus $m_{\rm F343N}-m_{\rm F814W}$ CMD, where ${\rm AGB}_{\rm II}$
stars are slightly redder then ${\rm AGB}_{\rm I}$ stars.  The maximum
difference between ${\rm AGB}_{\rm I}$ and ${\rm AGB}_{\rm II}$ is
$\sim 0.2$ in the $m_{\rm F275W}-m_{\rm F814W}$ CMD.

These diagrams demonstrate that the AGB of M\,15 hosts more than
  one stellar population. Spectroscopy is mandatory to connect
  multiple populations along the AGB and the RGB and to understand 
  whether
  the RGB stars with extreme chemical composition ascend the AGB phase 
 or avoid this evolutionary phase
  (\citealt{2013Natur.498..198C,2014A&A...571A..81C,2016ApJ...826L...1L,2017A&A...605A..98C,2017ApJ...843...66M}).

\section{Population ratios}
\label{sect:frac}
To determine the fraction of RGB stars in each of the five populations (POPs\,A--E),
  we extended to M\,15 the technique
  introduced by our group in the investigation of NGC\,2808
  (\citealt{2012A&A...537A..77M}; \citetalias{2016MNRAS.463..449S}).  At
  odds to what we have done in previous papers, which were based on
  the CMDs, here for the first time we exploit the ChM.

We first defined five regions R$_{\rm i}, i=1,...,5$, in the ChM as
shown in the left panel of Fig.~\ref{fig12}.  Each region is an
ellipse centred on each sub-population of M\,15 and is
similar to the ellipse that best reproduces the distribution of the
observational errors.

Since the stellar populations of M\,15 are partially overlapped in the
ChM, each region R$_{\rm i}$ would include stars from all the five
sub-populations. Specifically, the number of stars within each region
$N_{\rm i}$ can be expressed as:
 \begin{equation}
N_{\rm i} = N_{\rm A} f^{\rm A}_{\rm i}+N_{\rm B} f^{\rm B}_{\rm i}+N_{\rm C} f^{\rm C}_{\rm i}+N_{\rm D} f^{\rm D}_{\rm i}+N_{\rm E} f^{\rm E}_{\rm i}\,\,\,\,\,\,\,{\rm with}\,\,\,i=1,...,5
\label{eq:3}
\end{equation}
where $N_{\rm A}$-$N_{\rm E}$ are the numbers of POP\,A-POP\,E stars
in the ChM and $f^{\rm A-E}_{\rm i}$ are the fractions of POP\,-POP\,E
stars in the region R$_{\rm i}$ of the ChM.

To estimate the values of $f^{\rm A-E}_{\rm i}$, we simulated the ChM
of each stellar population of M\,15 by using ASs and the same
procedure described in the previous Section.  An example is provided
in the right panel of Fig.~\ref{fig12} where we highlight the simulated ChM
for POP\,A stars. The fractions of POP\,A stars in the five
regions of the ChM, $f^{\rm A}_{\rm i}$, are calculated as the ratio
between the number of ASs within each region and the total number of
simulated stars.  We used the same method to determine $f^{\rm
  B-E}_{\rm i}$.

To derive the number of stars in the ChM that belong to each
population, $N_{\rm A}$-$N_{\rm E}$, we solved the system of five
equations (\ref{eq:3}). We find that POP\,A, POP\,B, and POP\,C are
the most populous stellar populations in M\,15, making up 
(27$\pm$2)\%, (26$\pm$3)\%, and (29$\pm$3)\% of RGB stars,
respectively.  POP\,D contains (13$\pm$2)\% of RGB stars while POP\,E
is formed by (5$\pm$1)\% of RGB stars.

Finally, we verified that the final result is not significantly
affected either by the size of the regions $R_{\rm i}$ nor by the
exact location of their centres.  In Figure~\ref{fig12} we assumed
that the lengths of both semi-axes of the ellipses are 2.5 times
bigger than the corresponding dispersion expected for a single
population ($\sigma_{\rm obs}$).  We repeated the procedure by using
ellipses with axes that are 2.0 and 3.0 bigger than $\sigma_{\rm obs}$
and find that the fractions of POP\,A--E stars are the same within
2\,\%.  Similarly, we shifted the centres of each region by
$\pm\frac{1}{2}$semi-major axis along $\Delta_{\rm F275W,F814W}$, and
$\pm\frac{1}{2}$semi-minor axis along $\Delta_{C\,{\rm
    F275W,F336W,F438W}}$ and find that the derived fractions of
POP\,A--E stars remain unchanged within 2\,\%.

In  Appendix \ref{ap:1} we provide demonstrations of the
  reliability of the method that we used to derive the fraction of
  stars in each population.

\begin{figure}
\centering
\includegraphics[width=0.5\textwidth, bb=20 151 342 700]{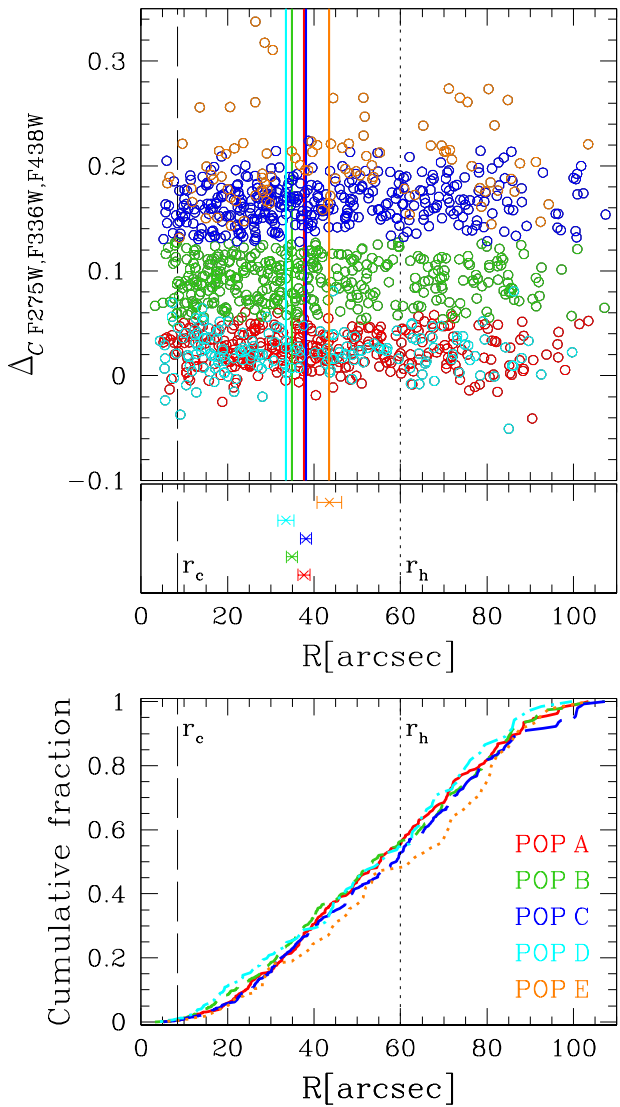}
\caption{Top panel shows the pseudo-colour $\Delta_{C {\rm
      F275W,F336W,F438W}}$ as a function of the distance from the
  centre of M\,15. The red, green, blue, cyan, and orange vertical
  lines are the median radial coordinates of POP\,A, POP\,B, POP\,C,
  POP\,D, and POP\,E, respectively. The median radial coordinates of
  the different populations are also reported in the middle panel,
  with the error-bars as shown. Bottom panel shows the cumulative
  radial distributions for each population. All populations
  share the same radial trend. Dashed and dotted vertical lines are
  the core (8.4\,arcsec) and the half-light (60\,arcsec) radii, as
  reported in \citet[updated to December
    2010]{1996AJ....112.1487H}. \label{fig13}}
\end{figure}

\section{The radial distribution of multiple populations}
\label{sec:raddis}

 To derive the radial distribution of the stellar populations in M\,15
  we adopted two different techniques.  We first calculated the median
  radial distance from the cluster centre of each population, $R_{\rm
    med, POP\,A-E}$, in close analogy with what has been done by
  \citet{2015ApJ...804...71L}. We expect that two populations with the
  same radial distribution would have the same value of $R_{\rm med}$,
  while different values of $R_{\rm med}$ would imply different radial
  distributions, with the more centrally-concentrated population
  having also smaller values of $R_{\rm med}$.

Results are illustrated in the upper panel of Fig.~\ref{fig13} where
we plot the $\Delta_{C {\rm F275W,F336W,F438W}}$ pseudo-colour of RGB
stars as a function of the radial distance from the cluster centre.

We find: $R_{\rm med, POP\,A}=(37.6\pm1.4)$\,arcsec (red line),
$R_{\rm med, POP\,B}=(34.9\pm1.2)$\,arcsec (green line), $R_{\rm med,
  POP\,C}=(38.1\pm1.3)$\,arcsec (blue line), $R_{\rm med,
  POP\,D}=(33.5\pm1.8)$\,arcsec (cyan line), $R_{\rm med,
  POP\,E}=(43.5\pm2.8)$\,arcsec (orange line), with errors estimated
via bootstrapping.

For completeness, we calculate the median radial distance for the
sample of POP\,A$+$D, $R_{\rm med, POP\,A+D}=(36.7\pm1.2)$\,arcsec,
and of POP\,C$+$E stars, $R_{\rm med, POP\,C+E}=(38.7\pm1.3)$\,arcsec.

All the measured values of $R_{\rm med, POP\,A-E}$ are consistent
within 2.2 $\sigma$ thus demonstrating that none of the populations of
M\,15 is significantly more centrally concentrated than the others
inside $\sim 2\times r_{\rm h}$.

In the lower panel of Fig.~\ref{fig13}, we compare the cumulative
radial distributions of the five populations. The Kolmogorov-Smirnov
test shows that all the populations are consistent with having the
same radial distrubions at the 95\% confidence level.

 To compute the cumulative radial distribution and the median
  radial distance of POPs A-E, we adopted the identification of the
  stellar populations illustrated in Sect.~\ref{sub:mpops} and in Fig.~\ref{fig6}. Computing
  the radial distributions using this approach could give weak
  results, because of the overlapping of the different populations in
  the various CMDs and ChM.

\begin{figure*}
\includegraphics[width=.9990\textwidth]{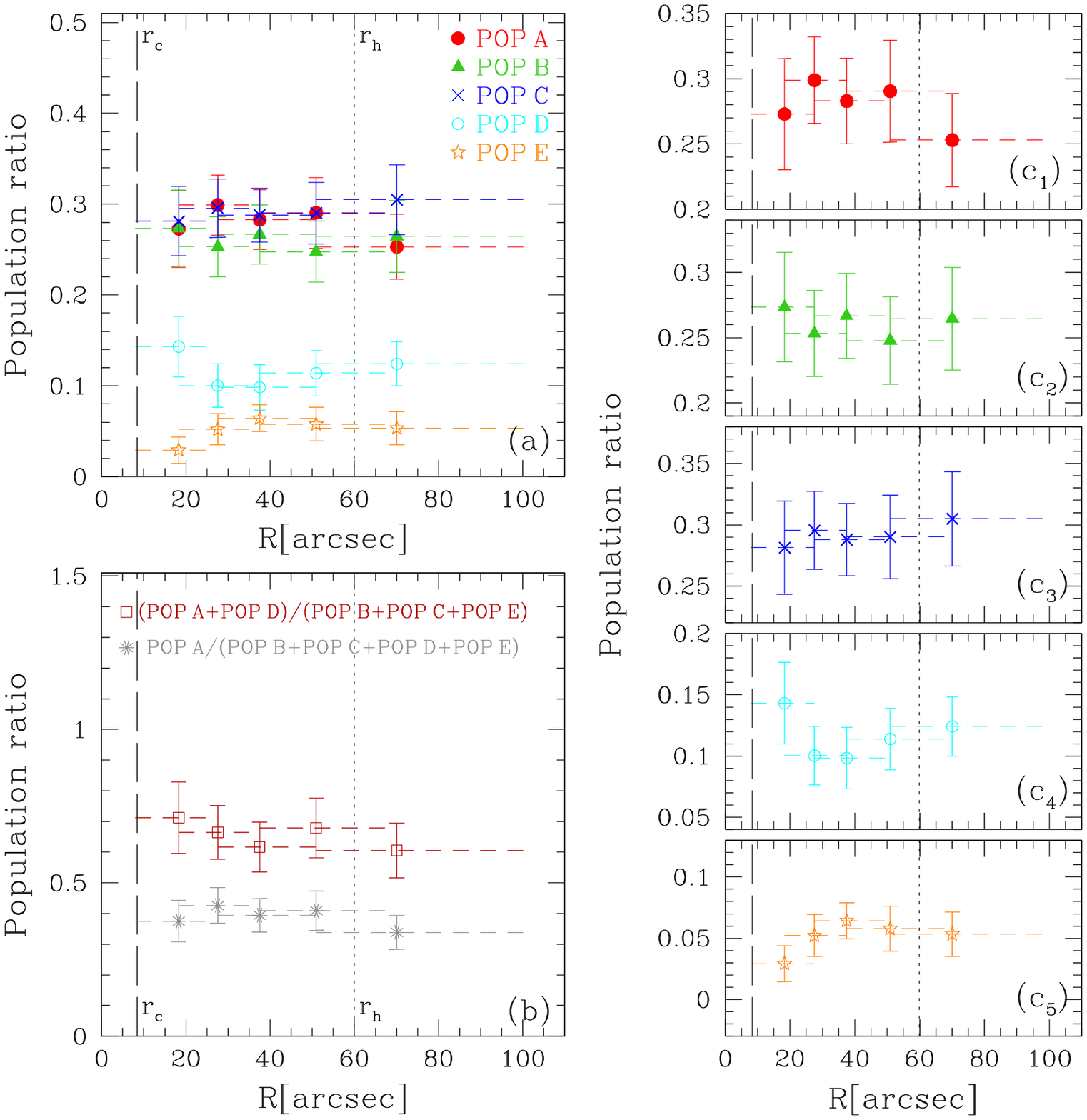}
\caption{The radial distributions of the five stellar populations
  hosted by M\,15 are shown in panel (a). Panel (b) shows the radial
  distribution of 1G/2G stars under two different hypotheses: (i) 1G is
  composed by only POP\,A stars (in grey); (ii) 1G is formed by POP\,A
  and POP\,D stars (in brown). Panels (c) are a zoom on the radial
  distribution of each single population. In all the panels, dashed
  and dotted vertical lines are the core and the half-light radii, as
  reported in \citet[updated to December
    2010]{1996AJ....112.1487H}. \label{fig14}}
\end{figure*}

\begin{table*}
  \caption{Fraction of POP\,A, POP\,B, POP\,C, POP\,D,  and POP\,E stars at different
    radial distances from  M\,15 centre.}  
  \begin{tabular}{ r r r c c c c c}
\hline
\hline
\multicolumn{1}{c}{$R_{\rm min}$} &
\multicolumn{1}{c}{$R_{\rm max}$} &
\multicolumn{1}{c}{$R_{\rm ave}$} &
\multicolumn{1}{c}{$f_{\rm POP\,A}$} &
\multicolumn{1}{c}{$f_{\rm POP\,B}$} &
\multicolumn{1}{c}{$f_{\rm POP\,C}$} &
\multicolumn{1}{c}{$f_{\rm POP\,D}$} &
\multicolumn{1}{c}{$f_{\rm POP\,E}$} \\
\multicolumn{1}{c}{[arcsec]} &
\multicolumn{1}{c}{[arcsec]} &
\multicolumn{1}{c}{[arcsec]} &
\multicolumn{1}{c}{  } &
\multicolumn{1}{c}{  } &
\multicolumn{1}{c}{  } &
\multicolumn{1}{c}{  } &
\multicolumn{1}{c}{  } \\
\hline
        7.95   &    27.51   &   18.31   &    0.27$\pm$0.04   &   0.27$\pm$0.04   &   0.28$\pm$0.04   &   0.15$\pm$0.03   &   0.03$\pm$0.01\\
       18.31   &    37.46   &   27.55   &    0.30$\pm$0.03   &   0.25$\pm$0.03   &   0.30$\pm$0.03   &   0.10$\pm$0.02   &   0.05$\pm$0.02\\
       27.55   &    50.93   &   37.52   &    0.28$\pm$0.03   &   0.27$\pm$0.03   &   0.29$\pm$0.03   &   0.10$\pm$0.03   &   0.06$\pm$0.02\\
       37.52   &    70.08   &   50.95   &    0.29$\pm$0.04   &   0.25$\pm$0.03   &   0.29$\pm$0.03   &   0.11$\pm$0.02   &   0.06$\pm$0.02\\
       50.95   &   101.37   &   70.10   &    0.25$\pm$0.04   &   0.27$\pm$0.04   &   0.31$\pm$0.04   &   0.12$\pm$0.02   &   0.05$\pm$0.02\\

\hline

\end{tabular}

%
%
%

  \label{tab2}
\end{table*}

As an alternative approach to deriving the radial distribution of MPs
in M\,15, we divided the field of view into five radial bins, each
containing the same number of RGB stars (422). We applied the
procedure described in Sect.~\ref{sect:frac} (that does not
  suffers of the problems due to the overlapping of the populations in
  the ChM) to the ChMs obtained with the stars in each radial bin, we
derived the fraction of POP\,A--E stars in each radial interval and we
used them to analyse the radial distribution of each population.

The radial distributions of the five stellar populations are shown in
panel (a) and zoomed in panels (c) of Fig.~\ref{fig14};  In
  Table~\ref{tab2} we list the fractions of POP\,A--E stars in the
  five analysed radials bins.

As shown in Figure~\ref{fig14} and Table~\ref{tab2}, the
distributions of all the populations are flat within the errors. The
slopes of the best-fit straight lines are:
$(-5.9\pm4.0)\cdot10^{-4}$\,arcsec$^{-1}$ (POP\,A);
$(-1.2\pm3.1)\cdot10^{-4}$\,arcsec$^{-1}$ (POP\,B); $(
3.1\pm1.7)\cdot10^{-4}$\,arcsec$^{-1}$ (POP\,C); $(
1.4\pm4.8)\cdot10^{-4}$\,arcsec$^{-1}$ (POP\,D); $(
4.2\pm3.5)\cdot10^{-4}$\,arcsec$^{-1}$ (POP\,E) and are consistent
with zero within $\sim$ 2 $\sigma$.

Finally, in panel (b) of Figure~\ref{fig14} we plot the radial
distribution of the ratio between POP\,A$+$POP\,D and
POP\,B$+$POP\,C$+$POP\,E stars (brown squares). These groups of stars
correspond to the 1G and 2G as defined in
\citetalias{2017MNRAS.464.3636M}. We also show the ratio between the
stars of POP\,A and of the remaining stellar populations as a function
of the radial distance from the cluster centre (grey asterisks).  The
slopes of the straight lines that best match the brown and grey points
are $(-11.0\pm7.7)\cdot10^{-4}$\,arcsec$^{-1}$ and
$(-13.5\pm9.6)\cdot10^{-4}$\,arcsec$^{-1}$, respectively. This demonstrates
that in both cases the observations are consistent with a flat
distribution within the errors.

\section{Summary and Conclusions}
\label{sec:summ}
 As part of the {\it HST} UV survey of Galactic GCs
 (\citetalias{2015AJ....149...91P}), we used multi-band {\it HST}
 photometry from GO-12605 and from the archive to investigate multiple
 populations in M\,15 and to analyse their radial distribution.

  Previous papers have shown that M\,15 hosts three main groups of RGB
  stars (\citetalias{2015AJ....149...91P};
  \citealt{2015ApJ...804...71L}). Our results increase the complexity
  of the multiple-population phenomenon in this GC.  Indeed, by using
  the ChM, we provide  evidence for spreads around POP\,A and POP\,C. These spreads are not compatible, within photometric errors, with the idea that such populations are composed by single stellar populations. An explanation of these spreads could be the presence of two additional populations.
  Specifically, the group of 1G stars, identified in
  \citetalias{2017MNRAS.464.3636M} includes two sub-populations,
  which we named A and D, while 2G stars include populations B, C, and E.
  
  We derive the fraction of stars within each population with respect
  to the total number of RGB stars and find that the three dominant
  populations, A, B, and C, include (27$\pm$2)\%, (26$\pm$3)\%, and
  (29$\pm$3)\% of stars, respectively.  POP\,D and POP\,E are less
  populated and host (13$\pm$2)\% and (5$\pm$1)\% of RGB stars,
  respectively.

  M\,15 exhibits a poorly-populated SGB that is seen to be fainter than the
  majority of SGB stars in all the CMDs we were able to analyze, including those
  CMDs, like $m_{\rm F814W}$ vs.\,$m_{\rm F606W}-m_{\rm F814W}$ and
  $m_{\rm F438W}$ vs.\,$m_{\rm F438W}-m_{\rm F606W}$, that are constructed 
  with optical filters only.  Moreover, the faint SGB of M\,15 seems to evolve
  into the RGB-E, which has redder $m_{\rm F336W}-m_{\rm F814W}$
  colours than the bulk of RGB stars with the same luminosity.

  In \citetalias{2017MNRAS.464.3636M} we found that about 18\%
  of the 57 analysed clusters, hereafter Type\,II GCs, exhibit a
  number of distinctive features.  These include a split SGB in
  optical CMDs with the faint SGB evolving into a red RGB in the
  $m_{\rm F336W}$ vs.\,$m_{\rm F336W}-m_{\rm F814W}$ CMD. Moreover,
  red-RGB stars define a distinct locus in the ChM.  Type\,II GCs
  have been widely investigated spectroscopically
  (e.g.\,\citealt{2015MNRAS.450..815M}, see their Table~10). In
  contrast with the majority of GCs, which are mono-metallic
  (e.g.\,\citealt{2009A&A...508..695C}), the red-RGB stars and the
  faint SGB of these clusters are enhanced in metallicity, C$+$N$+$O,
  and in $s$-process elements with respect to the remaining cluster
  stars
  (e.g.\,\citealt{2009A&A...505.1099M,2011A&A...532A...8M,2012A&A...541A..15M,
    2009ApJ...695L..62Y,2014MNRAS.441.3396Y,2015AJ....150...63J,2017ApJ...836..168J}).
  The photometric similarity between M\,15 and the Type-II GCs
  suggests that M\,15 belongs to this class of clusters.

However, it is worth noting that M\,15 shows no evidence for
internal metallicity variations
(\citealt{1999Ap&SS.265..145S,2009A&A...508..695C}) but hosts two
stellar groups with different abundances of Barium
(\citealt{1999Ap&SS.265..145S,2011AJ....141..175S,2013A&A...553A..47W}). 
Nevertheless, it is unlikely that the population of Ba-rich stars discovered by
Sneden and collaborators, which includes about half of the analysed
stars, corresponds to our POP\,E, which includes only $\sim$5\% of the
total number of stars.  Moreover, in contrast with other Type\,II GCs
where the production of Barium is attributed to s-processes, the
enhancement in Barium within M\,15 has been associated with r-processes
as indicated by the correlation between [Ba/Fe] and [Eu/Fe]
(\citealt{1999Ap&SS.265..145S}).  Spectroscopy of a large sample of
stars, particularly including POP\,E stars, is mandatory to establish whether M\,15
shares spectroscopic similarities with the other Type\,II GCs or not.

   One important result of this
  paper concerns the radial distribution of MPs in M\,15. The analysis
  of present-day radial distribution of 1G and 2G stars in GCs permits
  to discriminate between different formation scenarios of MPs in
  GCs.

Papers in the literature show that in some GCs, such as $\omega$\,Cen, 47\,Tuc,
and NGC\,2808, 2G stars are more centrally concentrated than the 1G stars
(\citealt{2007ApJ...654..915S,2009A&A...507.1393B,2012ApJ...744...58M,2018ApJ...853...86B};  \citealt{2011A&A...525A.114L};
\citetalias{2016MNRAS.463..449S}). In other clusters, like NGC\,6752
and M\,5, 1G and 2G stars share the same distribution, indicating that
the two populations are mixed due to dynamical evolution
(e.g.\,\citealt{2013ApJ...767..120M, 2015A&A...573A..70N,
  2017ApJ...844...77L}).

The possibility that 2G stars formed in the central regions of the
proto-GCs has been recently challenged by
\citet{2015ApJ...804...71L}. These authors have detected three stellar
populations along the RGB of M\,15 and found that the population with
primordial chemical composition is the most centrally concentrated.

In this paper, we have demonstrated that the stellar populations hosted by
this short-relaxation-time (  half-mass relaxation time $t(r_h)\sim 9.32$;  \citealt[updated to
  December 2010]{1996AJ....112.1487H}),
post-core-collapse system share
the same radial distribution, in contrast with previous
conclusions. Moreover we were able to verify that the group formed by POP\,A and
POP\,D stars, which approximately corresponds to the stellar
population with primordial chemical composition, has the same radial
distribution as second-generation stars.

\appendix

\begin{figure*}
\centering
\includegraphics[width=0.85\textwidth]{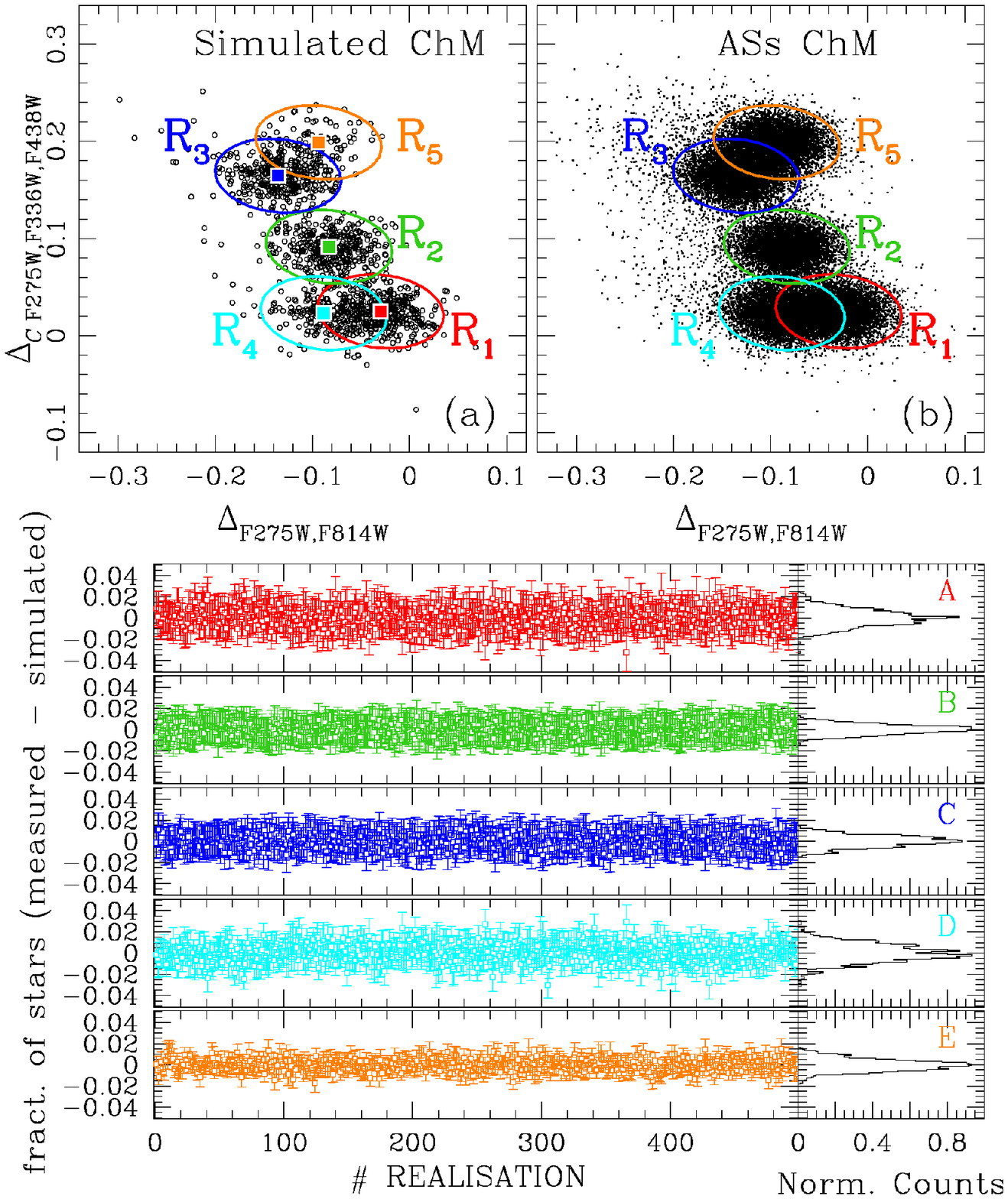}
\caption{Panels (a) and (b) show the simulated and the ASs ChMs. The
  coloured eclipses are the regions R$_i$ within our analysis has been
  performed. Left-hand bottom panels shows the difference between the
  measured and the expected fraction of stars at each of the 500
  realisations of the simulated ChM. As shown in the right-hand bottom
  panels, the distributions of these differences is centred on
  zero.  \label{figA1}}
\end{figure*}

\section {Reliability of the measured fraction of MP stars}
\label{ap:1}
To test the reliability of the method adopted in
Sect.~\ref{sect:frac}, we built 500 synthetic $\Delta_{C\,{\rm
    F275W,F336W,F438W } } $ versus $\Delta_{\rm F275W,F814W}$ ChMs
with five different populations. Each synthetic population is obtained
as follows.  We first constructed the ChM of a single population using the
ASs, as explained in Sect.~\ref{sect:frac}.  From this sample of stars
we randomly extracted $N_{\rm synth} + \Delta N_{\rm synth}$ stars,
where $\Delta N_{\rm synth}$ is a random number between -50 and 50 and
$N_{\rm synth}$ is a number different for each simulated population,
in such a way the final number of stars in the simulated ChM is
similar to the number of stars in the observed ChM. In particular we
adopted $N_{\rm synth}=(350,350,350,150,60)$ for POP\,(A,B,C,D,E),
respectively. We centred each synthetic population in the centres of
the observed population. Panel (a) of Fig.~\ref{figA1} show a
realisation of the synthetic ChM. Panel (b) is the ASs ChM used to
estimate the contamination of each population X in the region R$_{\rm
  Y}$ centred and containing the bulk of the population Y. We
measured the fraction of stars that belong to each population using
the technique illustrated in Sect.~\ref{sect:frac}. The bottom panel shows
the comparison between the fraction of stars measured from the
synthetic ChM and the fraction of stars expected for each of the 500
realisations and for each of the simulated population.  The error bars
are obtained as explained in Sect.~\ref{sect:frac}, and are a
combination of Poissonian error, uncertainty on the centre and size of
the adopted regions; inside the errors the measured fraction of stars
reproduce the simulated one.

The right-hand panels show the distribution of the difference between the
fraction of stars measured and the fraction of stars expected: the
distributions shown in the right-bottom panels may be approximated to
gaussian distributions centred on 0 and with a standard deviation
$\lesssim 0.01$, confirming that the method adopted in
Sect.~\ref{sect:frac} to measure the fraction of stars in each
population is reliable.

\section*{Acknowledgements}

We thank the anonymous referee for the useful comments and suggestions.
D.N. and G.P. acknowledge support by the Universit\`a degli Studi di
Padova Progetto di Ateneo CPDA141214, ``Toward Understanding Complex
Star Formation in Galactic Globular Clusters''. G.P. and
M.L. acknowledge support by PRIN-INAF2014, ``The Kaleidoscope of
Stellar Populations in Galactic Globular Clusters with Hubble Space
Telescope''.  A.P.M.\ acknowledges support by the European Research
Council through the ERC-StG 2016 project 716082 `GALFOR'.A.F.M. has been
supported by the Australian Research Council through Discovery Early
Career Researcher Award DE160100851.  J.A.\ acknowledges the support
of STScI grant GO-13297.


\bibliographystyle{mnras}
\bibliography{biblio} 

\bsp	
\label{lastpage}
\end{document}